\documentclass[useAMS,usenatbib]{mn2e}
\usepackage{times}
\usepackage{graphicx, amssymb}

\newcommand\nodata{ ~$\cdots$~ }
\newcommand{\oi}{O$\;${\small\rm I}\relax}
\newcommand{\oii}{O$\;${\small\rm II}\relax}
\newcommand{\oiii}{O$\;${\small\rm III}\relax}
\newcommand{\oiv}{O$\;${\small\rm IV}\relax}
\newcommand{\ov}{O$\;${\small\rm V}\relax}
\newcommand{\ovi}{O$\;${\small\rm VI}\relax}
\newcommand{\ovii}{O$\;${\small\rm VII}\relax}
\newcommand{\oviii}{O$\;${\small\rm VIII}\relax}
\newcommand{\cii}{C$\;${\small\rm II}\relax}
\newcommand{\ciii}{C$\;${\small\rm III}\relax}
\newcommand{\civ}{C$\;${\small\rm IV}\relax}
\newcommand{\nii}{N$\;${\small\rm II}\relax}
\newcommand{\niii}{N$\;${\small\rm III}\relax}
\newcommand{\niv}{N$\;${\small\rm IV}\relax}
\newcommand{\nv}{N$\;${\small\rm V}\relax}
\newcommand{\svi}{S$\;${\small\rm VI}\relax}
\newcommand{\neviii}{Ne$\;${\small\rm VIII}\relax}
\newcommand{\mgx}{Mg$\;${\small\rm X}\relax}
\newcommand{\sixii}{Si$\;${\small\rm XII}\relax}
\newcommand{\feii}{Fe$\;${\small\rm II}\relax}
\newcommand{\siii}{Si$\;${\small\rm II}\relax}
\newcommand{\siiii}{Si$\;${\small\rm III}\relax}
\newcommand{\siiv}{Si$\;${\small\rm IV}\relax}
\newcommand{\bvalues}{$b$-values}
\newcommand{\lya}{Ly$\alpha$}
\newcommand{\lyb}{Ly$\beta$}
\newcommand{\lyg}{Ly$\gamma$}
\newcommand{\HI}{H$\;${\small\rm I}\relax}

\newcommand{\heii}{He$\;${\small\rm II}\relax}
\newcommand{\kms}{km~s$^{-1}$\relax}
\newcommand{\percc}{cm$^{-3}$\relax}
\newcommand{\nav}{$N_a(v)$}
\newcommand{\e}[1]{10^{#1}}
\newcommand{\z}{$z$}
\newcommand{\fuse}{{\em FUSE}}
\newcommand{\hst}{{\em HST}}
\newcommand{\pks}{PKS~0405--123}
\newcommand{\dave}{Dav\'{e}}
\newcommand{\fraccoll}[1]{\ensuremath{\phi_{\rm coll}(\mbox{#1})}}

\title[$z\sim0.5$ O\,{\sevensize\it VI} Absorption]{Strong $z\sim0.5$
  O\,{\sevensize\bf VI} Absorption Toward PKS~0405-123: Implications
  for Ionization and Metallicity of the Cosmic Web\thanks{Based on
    observations made with the NASA-CNES-CSA Far Ultraviolet
    Spectroscopic Explorer. FUSE is operated for NASA by the Johns
    Hopkins University under NASA contract NAS5-32985.  Also based on
    observations made with the NASA/ESA Hubble Space Telescope,
    obtained from the Data Archive at the Space Telescope Science
    Institute, which is operated by the Association of Universities
    for Research in Astronomy, Inc., under NASA contract NAS
    5-26555. These observations are associated with program \#7576.}}

\author[J.C. Howk et al.]{J. Christopher Howk,$^2$ Joseph
  S. Ribaudo,$^2$ Nicolas Lehner,$^2$
  \newauthor J. Xavier Prochaska,$^3$ Hsiao-Wen Chen$^4$\\
  $^2$Department of Physics, University of Notre Dame, Notre Dame, IN, USA \\
  $^3$ UCO/Lick Observatory; University of California, Santa Cruz;
  Santa Cruz, CA, USA\\
  $^4$Department of Astronomy \& Astrophysics, University of Chicago,
  Chicago, IL 60637}

\date{Accepted XXX.
      Received XXX.}

\pagerange{\pageref{firstpage}--\pageref{lastpage}}
\pubyear{2009}

\begin{document}

\maketitle

\label{firstpage}


\begin{abstract}

  We present observations of the intervening \ovi\ absorption line
  system at $z_{abs}=0.495096$ toward the QSO \pks\ ($z_{em} =
  0.5726$) obtained with the {\em Far Ultraviolet Spectroscopic
    Explorer} and Space Telescope Imaging Spectrograph on board the
  {\em Hubble Space Telescope}.  In addition to strong \ovi, with
  $\log N(\mbox{\ovi}) = 14.47\pm0.02$, and moderate \HI, with $\log
  N(\mbox{\HI}) = 14.29\pm0.10$, this absorber shows absorption from
  \ciii, \niv, \oiv, and \ov, with upper limits for another seven
  ions. The large number of available ions allows us to test
  ionization models usually adopted with far fewer contraints.  We
  find that the observed ionic column densities cannot be matched by
  single temperature collisional ionization models, in or out of
  equilibrium.  Photoionization models can match all of the observed
  column densities, including \ovi.  If one assumes photoionization by
  a UV background dominated by QSOs, the metallicity of the gas is
  [O/H]$\, \approx-0.15$, while if one assumes a model for the UV
  background with contributions from ionizing photons escaping from
  galaxies the metallicity is [O/H]$\, \approx-0.62$.  Both give
  [N/O]$\, \sim-0.6$ and [C/H]$\, \sim-0.2$ to $\sim-0.1$, though a
  solar C/O ratio is not ruled out.  The choice of ionizing spectrum
  is poorly constrained and leads to systematic abundance
  uncertainties of $\approx0.5$ dex, despite the wide range of
  available ions.  Multiphase models with a contribution from both
  photoionized gas (at $T\sim10^4$ K) and collisionally ionized gas
  (at $T\sim(1-3)\times10^5$ K) can also match the observations for
  either assumed UV background giving very similar metallicities. The
  \ovi\ in this system is not necessarily a reliable tracer of WHIM
  matter given the ambiguity in its origins.  We do not detect
  \neviii\ or \mgx\ absorption.  The limit on \neviii / \ovi$\,<0.21$
  ($3\sigma$), the lowest yet observed.  Thus this absorber shows no
  firm evidence of the ``warm-hot intergalactic medium'' at $T \sim
  (0.5-3)\times 10^6$ K thought to contain a significant fraction of
  the baryons at low redshift.  We present limits on the total column
  of warm-hot gas in this absorber as a function of temperature.  This
  system would be unlikely to provide detectable X-ray absorption in
  the ions \ovii\ or \oviii\ even if it resided in front of the
  brighter X-ray sources in the sky.

\end{abstract}



\section{Introduction}
\label{sec:intro}

The baryon density of the universe, $\Omega_b$, is a fundamental
quantity in cosmology, and identifying the location and physical state
of the baryons is a focus of significant work (see Prochaska \&
Tumlinson 2008 for a recent review).  A census of luminous, emitting
baryons in the universe falls short of the total baryon density of the
universe determined from other means.  Various methods of calculating
the baryon density are in good agreement, including the comparison of
primordial deuterium measurements with Big Bang nucleosynthesis (e.g.,
O'Meara et al. 2006, Kirkman et al. 2003), and cosmic microwave
background studies with the new WMAP measurements (e.g., Spergel et
al. 2006).  At high redshift, baryonic matter traced in emission falls
short of the total baryon budget because most of the baryons reside
and can be found within the cool, photoionized intergalactic medium
(IGM) of the \lya\ forest (LAF; e.g., Rauch et al. 1997).  At lower
redshift, $z\la0.5$, the census of stars and gas in galaxies, galaxies
and gas in clusters, etc. still falls short of $\Omega_b$, as
emphasized by a number of researchers (Persic \& Salucci 1992;
Fukugita, Hogan, \& Peebles 1998).  This baryon deficit persists even
if the cool, photoionized LAF is included.  The emitting components
contribute only $\sim0.1\Omega_b$ (Prochaska \& Tumlinson 2008), while
including the LAF may increase this to $\sim0.3\Omega_b$ (e.g.,
Danforth \& Shull 2008, Lehner et al. 2007, Penton, Stocke, \& Shull
2004), though with significant uncertainty.

Thus, the easy to identify baryonic constituents of the low-redshift
universe represent $\la50\%$ of the total baryon budget.  Fukugita et
al. (1998) suggested that hot gas in the halos of galaxies and in
groups of galaxies might help solve this ``missing baryons problem.''
An additional component of the baryon budget was suggested by
numerical cosmological simulations, in which $\Omega_b$ is an initial
condition.  Such efforts (e.g., Cen \& Ostriker 1999, \dave\ et
al. 2001) pointed out that the mass scale of structures in the
universe grows with time, and gas falling into these structures at low
redshift might then be heated to quite high temperatures.  These works
suggested that a significant fraction of the low-redshift baryons
might then be found in a ``warm-hot intergalactic medium'' (WHIM) at
tempertures $T \approx 10^{5-7}$ K and densities peaked at $\sim10$
times the mean density (e.g., Cen \& Ostriker 2006).  A comparison of
a range of models suggests the WHIM is a robust prediction of modern
simulations, which predict $30\% - 50\%$ of the low-redshift baryons
might be found in this phase of the IGM (Cen \& Ostriker 2006, \dave\
et al. 2001).  More recent numerical and analytical works have
supported this general conclusion while exploring the effects of
feedback (e.g., from galactic winds), non-equilibrium ionization, and
cooling on these predictions and their connection to observables
(e.g., Oppenheimer \& \dave\ 2009, Cen \& Ostriker 2006, Cen \& Fang
2006, Kang et al. 2005, Furlanetto et al. 2005, Fang \& Bryan 2001,
Phillips et al. 2001, Cen et al. 2001).

Given the low densities and high temperatures predicted for the WHIM,
the gas in this phase is difficult to detect in emission or
absorption.  One approach to studying the baryon content of the WHIM
is to identify broad \lya\ absorption in the low-\z\ IGM.  Tripp et
al. (2001), Richter et al. (2004), Sembach et al. (2004), and Lehner
et al. (2007) report on the detection of broad \lya\ absorbers (BLAs)
having $b \ga 40$ \kms, which corresponds to $T\ga 10^5$ K if thermal
broadening predominates.  Lehner et al. (2007) conclude the fraction
of LAF clouds categorized as BLAs is higher at low-\z\ than at larger
redshifts (from the sample of Kirkman \& Tytler 1997), consistent with
predictions of cosmological simulations if thermal motions dominate
the broadening (see \dave\ et al. 2001).  Some of the BLAs may be
blends of narrow lines (or even small ripples in the observed QSO
continuum), a contamination that is difficult to assess with the
generally low signal-to-noise data available with STIS observations.
In one case, Richter et al. (2004) find \ovi\ absorption associated
with a BLA for which the ratio of \ovi\ to \HI\ $b$-values is
consistent with pure thermal broadening, strongly suggesting an origin
within hot gas at $T\sim3\times10^5$ K and supporting the idea that
some BLAs are thermally broadened.  Similarly, Tripp et al. (2001)
identified a blended, broad \lya\ line with associated \ovi\
absorption; a joint analysis assuming the \HI\ and \ovi\ arise from
the same gas yields $T\sim2\times10^5$ K.

In addition to broad \HI\ absorption, absorption lines from
highly-ionized metals may be useful for probing WHIM gas at $T\ga10^5$
K in filaments or groups (e.g., Verner et al. 1994b, Mulchaey et
al. 1996, Cen \& Ostriker 1999).  The highly ionized stages of oxygen,
the most abundant metal, are the best studied, including \ovi, \ovii,
and \oviii.  \ovii\ and \oviii\ have transitions accessible to X-ray
instruments, as does \ovi.  Searches for such absorbers have been
hampered by the relatively poor resolution of modern X-ray instruments
and the small number of sources observable at high enough signal to
noise by the {\em Chandra} and {\em XMM-Newton} observatories (see
recent summaries by Bregman 2007 and Richter et al. 2008).  While a
number of measurements have been presented in the literature (e.g.,
Nicastro et al. 2005), some results are questioned by other analyses
(Rasmussen et al. 2007, Kaastra et al. 2006) while a few may be on
firmer ground (e.g., Fang et al. 2007).  It is thus far difficult to
judge the fidelity of the X-ray measurements.

Given the high resolution and sensitivity of the currently-available
ultraviolet spectrographs, searches for absorption due to the \ovi\
$1031,1037$ \AA\ doublet may represent the most efficient means of
probing the WHIM, even though \ovi\ is not the dominant ionization
state of oxygen for the temperature at which the majority of the WHIM
is thought to exist ($\sim10^6$ K; \dave\ et al. 2001).  The first
blind search for low-redshift \ovi\ absorbers (mostly over $0.6 \la z
\la 1.3$) of Burles \& Tytler (1996) was only sensitive to very strong
absorbers but nonetheless demonstrated that these absorbers could
trace a significant component of the baryonic mass in the universe.
Tripp and collaborators first identified a number of \ovi\ absorption
line systems at very low redshift ($z<0.3$) with the {\em Far
  Ultraviolet Spectroscopic Explorer} ({\em FUSE}) and the Space
Telescope Imaging Spectrograph (STIS) on board the {\em Hubble Space
  Telescope} ({\em HST}); though their statistical sample was small,
these observations suggested a large number density of \ovi\ absorbers
at low-\z\ (Tripp, Savage, \& Jenkins 2000; Tripp \& Savage 2000).
Recent surveys of \ovi\ absorption at low redshift using a complete
sample of STIS observations continue to find a very large population
of absorbers (Tripp et al. 2008, Thom \& Chen 2008a, Danforth \& Shull
2008).  The number density of intervening \ovi\ absorbers per unit
redshift is $dN_{\rm O\,VI}/dz \approx 10 - 20$ for equivalent widths
$W_r (\lambda1031) > 30$ m\AA, perhaps increasing to $dN_{\rm
  O\,VI}/dz \sim 40$ for $W_r (\lambda1031) > 10$ m\AA\ (Danforth \&
Shull 2008).  For comparison, recent estimates for the number density
of \HI\ absorbers at $z<0.4$ give $dN_{\rm H\,I}/dz \approx 95$ for
$W_r (\lambda1215) > 90$ m\AA\ (Lehner et al. 2007) and $\approx130$
for $W_r (\lambda1215) > 30$ m\AA\ (Danforth \& Shull 2008).  The
large number of absorbers by itself suggests that the gas probed by
these \ovi\ transitions could indeed be an important reservoir of
baryons, although accurate estimates of the ionization state and
metallicity of these absorbers are required for a quantitative measure
of the total mass density.
 
However, it is possible or even likely that a significant fraction of
\ovi\ absorbers do not trace the WHIM, but rather probe
metal-enriched, photoionized IGM material.  This point has been
emphasized in a number of analyses of individual absorbers and sight
lines (Savage et al. 2002, Prochaska et al. 2004, Lehner et al. 2006,
Tripp et al. 2006, Cooksey et al. 2008), as well as in recent surveys
(Tripp et al. 2008, Thom \& Chen 2008b).  Oppenheimer \& \dave\ (2009)
have recently presented cosmological simulations in which \ovi\
absorbers predominantly trace gas at temperatures $\log T \approx
4.2$, well below the majority of the gas associated with the WHIM.  A
significant amount of the most commonly-used probe of WHIM gas, \ovi,
may therefore arise in photoionized gas, although Kang et al. (2005)
and Oppenheimer \& \dave\ (2009) have argued that some of the
photoionized \ovi\ may be WHIM gas that has cooled to low
temperatures.  It remains to be seen, however, if the relatively
extreme predictions of Oppenheimer \& \dave\ (2009) will be verified
by other simulations.

Undoubtedly some \ovi\ systems are associated with photoionized gas,
though whether they outnumber or dominate the WHIM absorbers is a
question open to debate (e.g., Danforth 2009).  Even if photoionized
absorbers represent a significant fraction of the total, most
simulations suggest that the highest equivalent width \ovi\ systems
are most likely to be collisionally ionized, either tracing WHIM gas
or gas associated with the hot halos of galaxies (Cen et al. 2001,
Fang \& Bryan 2001, Ganguly et al. 2008, Oppenheimer \& \dave\ 2009).

While \ovi\ may not be a perfect tracer of the WHIM, and X-ray
measurements are not yet secure, observations of EUV transitions that
are redshifted into the bands of \fuse\ and \hst\ provide other
evidence for shock-heated hot gas in the low-\z\ IGM.  In particular,
the doublets of \neviii\ $\lambda\lambda$770.409, 780.324, \mgx\
$\lambda\lambda$609.790, 624.950, and \sixii\ $\lambda\lambda$499.406,
520.665 are potentially useful probes of hot gas in redshifted IGM
(e.g., Verner, Tytler, \& Barthel 1994b\footnote{Verner et al. 1994b
  quote give wavelengths for the \mgx\ 624.950 \AA\ and \sixii\
  520.665 \AA\ transitions that are not in agreement with the
  wavelengths given in Verner, Barthel, \& Tytler 1994a.  The latter
  agree with those available via NIST.  We quote the NIST wavelengths
  throughout for these transitions.}) or intragroup (Mulchaey et
al. 1996) material; these ions peak in abundance in collisional
ionization equilibrium at temperatures of $T\approx6\times10^5,\
1.2\times10^6$, and $2\times10^6$ K, respectively (Gnat \& Sternberg
2007).  Given the large energies required to produce these ions, they
should represent good probes of hot, collisionally ionized gas.  They
probe temperatures in the same range as X-ray absorption lines, but at
better sensitivity to a given H column density given the maturity of
the UV instruments with which they can be detected.  Savage et
al. (2005) have reported on a detection of intervening \neviii\ at
$z\approx 0.21$ that they show comes from hot gas.  The \neviii
-bearing gas is part of a multiphase\footnote{The term multiphase will
  be used here to denote an absorber with a mixture of gas with
  differing temperatures and ionization states where these differences
  are often driven by differing ionization mechanisms.}  absorber with
strong \ovi\ in addition to complex low-ion absorption.  The ratio of
integrated column densities in this absorber is
$N(\mbox{\neviii})/N(\mbox{\ovi}) = 0.33\pm0.10$, which is consistent
with gas in colllisional ionization equilibrium near
$T\approx5\times10^5$ K.  Searches for \neviii\ absorption associated
with 10 other \ovi\ absorbers have thus far yielded no further
detections (Prochaska et al. 2004, Richter et al. 2004, Lehner et
al. 2006).  However, most of the \ovi\ absorbers are much weaker than
that discussed by Savage et al., and in only three of these
non-detections is the limit $N(\mbox{\neviii})/N(\mbox{\ovi}) < 0.5$.
Thus, \neviii\ has not yet been the subject of a comprehensive search
(see limits on $dN/dz$ in Prochaska et al. 2004).  To date, however,
most \neviii\ searches have used \ovi -based searches; if the two ions
do not coexist because the WHIM is quite hot, this may not be the best
approach.

Here we present a detailed analysis of the ionization state and
chemical abundances of the strong \ovi\ absorber at $z=0.495096$
toward PKS~0405-123.\footnote{Hereafter we simply refer to this as the
  $z=0.495$ absorber.}  The system was identified previously by
Bahcall et al. (1993) in \lya\ absorption using the Faint Object
Spectrograph on board {\em HST}.  Aspects of this absorber have
previously been discussed by Prochaska et al. (2004), Williger et
al. (2006), Lehner et al. (2006), Tripp et al. (2008), and Thom \&
Chen (2008a,b) based on \fuse\ and \hst\ observations.  This absorber
exhibits numerous metal-line absorption features in high-resolution
spectra obtained using {\em FUSE} and STIS, as previously mentioned by
Prochaska et al. (2004).  The large spectral coverage of the combined
{\em FUSE} and STIS spectra provides access to $\sim 15$ ions,
including six oxygen ions (O$^0$ through O$^{+5}$), and the
highly-ionized tracers of hot gas \neviii\ and \mgx. Coverage of the
large number of ionization states of oxygen and other elements gives
us the opportunity to study in detail the ionization mechanisms that
may be at work in this gas and place constraints on the temperature of
the absorbing material.  This absorber is among the strongest 5\%\ to
10\%\ of all \ovi\ systems, and the simulations predict strong
absorbers like this are more likely to be associated with hot,
collisionally-ionized material.  The results of our analysis show that
the $z\sim0.495$ absorber toward PKS~0405-123 must contain a
substantial photoionized component, and may or may not also contain
$\ga2\times10^5$ K gas as part of a multiphase structure.  The lack of
significant \neviii\ and \mgx\ places limits on the amount of hot WHIM
gas in this system.

We discuss out the observations used here in \S
\ref{sec:observations}.  In \S \ref{sec:absorber} we discuss our
measurements of the properties of the absorber.  We provide a detailed
analysis of the possible ionization mechanisms in this system in \S
\ref{sec:ionization}, and we discuss the implications in \S
\ref{sec:discussion}.  Finally, we summarize our principal results in
\S \ref{sec:summary}.


\section{Observations}
\label{sec:observations}

In this work we present observations of the \z = 0.495 absorber toward
\pks\ taken by several UV spectrographs.  Our analysis focuses on
three spectroscopic datasets at UV wavelengths: (1) FUV spectra
obtained with \fuse\ covering $\lambda \approx 912 - 1150$ \AA\ ($R
\approx 15000$); (2) STIS/E140M echelle spectroscopy acquired with
{\em HST} covering $\lambda \approx 1150 - 1700$ \AA\ ($R \approx
46000$); and (3) FOS/G190H spectroscopy from \hst\ covering $\lambda
\approx 1570 - 2330$ \AA\ ($R \approx 1300$).  We describe each of
these datasets briefly below.

\subsection{FOS Spectroscopy}

\pks\ was observed with the FOS using the G190H grating and the C-2
($0\farcs25\times2\farcs0$) aperture to feed the ``blue'' digicon as
part of GTO program 1025 (PI: Bahcall).  The total exposure time with
the FOS was 3.8 ksec.  These are pre-COSTAR data and have been
described by Bahcall et al. (1993) and Jannuzi et al. (1998), among
others.  We adopt the reduced data from the uniform FOS reduction of
Bechtold et al. (2002)\footnote{The data are available through {\tt
    http://lithops.as.arizona.edu/$\sim$jill/QuasarSpectra/}.} who
describe their reductions in detail.  We use the FOS data to study the
\lya\ absorption from the \z = 0.495 absorber.  This absorber is
unresolved at the resolution of FOS ($R\approx1300$ or $\Delta v
\approx230$ \kms\ FWHM).  The data have a signal-to-noise ratio $S/N
\approx 13$ per pixel, with four pixels per resolution element.

\subsection{STIS Spectroscopy}

\pks\ was observed by STIS using the E140M grating and the
$0\farcs2\times0\farcs06$ aperture to feed the FUV MAMA detector as
part of the GTO program 7576 (PI: Heap).  This setup yields spectra
with a velocity resolution $\Delta v \approx 7$ \kms\ (FWHM) with
approximately two pixels per resolution element.  The spectral
coverage of the STIS data is $\lambda \approx 1150 - 1700$ \AA .  The
total exposure time of the observations is 27.2 ksec.  The data have
$S/N\approx 5-7$ per pixel.  The absolute velocities are accurate to
$\sim0.5-1.0$ pixels ($\sim2- 4$ \kms) while the relative wavelength
scale is twice as good (Kim Quijano et al. 2003), though Tripp et
al. (2005) have noted occasional wavelength scale errors in excess of
these values.

The STIS data have been discussed extensively in the context of IGM
absorption by Chen \& Prochaska (2000) and Prochaska et al. (2004),
who discuss metal line absorbers, as well as Williger et al. (2006)
and Lehner et al. (2007), who concentrate on the properties of the
\HI\ absorption.  Our reduction is that used by Lehner et al. (2007).
Our measurements of the \z = 0.495 system will differ slightly from
Prochaska et al. given the slightly different reduction, our choice of
integration limits, reconsideration of continuum placement, and other
details.

\subsection{FUSE Spectroscopy}

\pks\ has been observed by \fuse\ (Moos et al. 2000; Sahnow et
al. 2000) as part of GI programs B087 (PI: Prochaska) and D103 (PI:
Howk) for a total exposure time of 142 ksec.  Prochaska et al. (2004)
and Chen \& Prochaska (2000) have reported on aspects of the IGM; the
data used here approximately double the exposure time of those earlier
observations.  The \fuse\ observations were taken in TTAG or photon
event mode with the object centred in the LWRS ($30\arcsec \times
30\arcsec$) apertures.  The \fuse\ data have been uniformly reduced
with CalFUSE v3.1 (Dixon et al. 2007).  These data have a
velocity resolution $\Delta v \sim 20$ \kms\ (FWHM) and are binned to
0.027 \AA\ per pixel ($\approx 7-8$ \kms\ per pixel), giving roughly 3
pixels per resolution element.  We have combined the data from all
channels in regions of overlap to provide S/N$\, \sim 10$ per pixel
over the spectral range covered by the LiF channels ($\lambda \sim
1000 - 1180$ \AA) and S/N$\, \sim 6$ per pixel over the range covered
by the SiC channels ($\lambda \sim 915 - 1000$ \AA).  While this
coaddition of channels can lead to a slight degradation in the breadth
and shape of the line spread function, the increase in S/N of data
outweighs this slight effect.

The absolute wavelength calibration of \fuse\ is not well determined.
We have set the zero point by comparing absorption lines from the
Galactic interstellar medium in the \fuse\ bandpass with similar lines
in the STIS bandpass, (e.g., comparing lines of \feii, \siii, and \oi\
between the instruments).  This approach generally leads to zero-point
uncertainties of order $\sim2-5$ \kms, although our experience has
shown this can occasionally be as large as 10 \kms.

\section{The $z=0.495$ Absorption System}
\label{sec:absorber}

The focus of this paper is the $z\approx0.495$ absorption system
toward \pks\ ($z_{em} = 0.574$).  Here we discuss the velocity
structure of the absorber and separately describe our analysis of the
metals and \HI\ in this system. 

\begin{figure}
\includegraphics[width = 8.truecm]{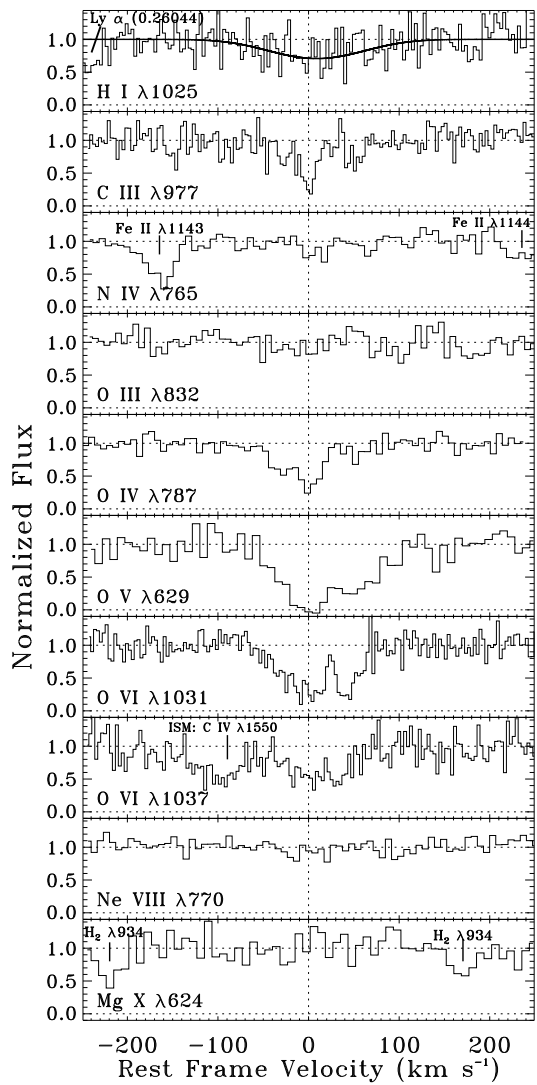}
\caption{Absorption profiles the $z=0.495096$ absorber toward
  PKS~0405--123 (the velocity zero-point was arbitrarily defined to
  match the central velocity of the main \ciii\ component derived
  through component fitting).  These data, ordered by atomic mass and
  ionization state of the ions, are from \fuse\ ($\lambda_0<790$ \AA)
  and STIS ($\lambda_0>790$ \AA). The detection of \ov\ is the first
  for intervening material at $z<1$.  The STIS data have a resolution
  of $\approx7$ \kms\ (FWHM) with two pixels per resolution element;
  the FUSE data have a resolution of $\approx20$ \kms\ with three
  pixels per resolution element. }
\label{fig:metalprofiles}
\end{figure}

Figure~\ref{fig:metalprofiles} presents the normalized profiles of
several important metal ions and \lyb\ (see \S \ref{sec:hydrogen}) in
this absorber for an assumed zero point of $z=0.495096$.  The redshift
is based on the centroid of the strong \ciii~977 \AA\ transition.  The
continua were estimated by fitting low-order Legendre polynomials to
regions free of absorption lines following Sembach \& Savage (1992).
Due to the simplicity of the quasar spectrum, most continua for this
work were first order (linear) fits.

We identify significant metal line absorption in the \ciii~977 \AA,
\niv~765 \AA, \oiv~787 \AA, \ov~629 \AA, and \ovi~1031, 1037 \AA\
transitions.  The measurement of \ov\ $\lambda 629$ is the first
direct detection of this ion in intervening absorbers at $z<1$ (though
see Telfer et al. 2002 for an indirect detection of the composite
absorption from many \lya\ forest clouds).  In addition, our data
include non-detections of strong lines from \cii, \nii, \niii, \oii,
\oiii, \neviii, and \mgx, among others.  We do not show profiles for
all the ions covered by our data, as the upper limits for several are
not particularly stringent or important in our final analysis.

\subsection{Metal Ion Absorption}
\label{sec:metals}

\subsubsection{Apparent Column Densities and Velocity Profiles}

Our measurements of the properties of the $z\approx0.495$ absorbtion
line system are presented in Table \ref{tab:totalcolumns}, including
our assumed atomic data [taken from the compilations of Verner et
al. (1994) and Morton (2003)].  The columns integrated over the
profiles are given, derived using a combination of the apparent
optical depth method (AOD) of Savage \& Sembach (1991) and profile
fitting techniques (described below).  The apparent optical depth,
$\tau_a$ of an absorption profile is written $\tau_a = -\ln \left[
  I(v)/I_c (v) \right]$, where $I(v)$ and $I_c(v)$ are the observed
intensity across the profile and the estimated continuum intensity,
respectively.  The apparent column density, \nav, is then $N_a(v) =
[(m_e c)\tau_a(v)]/[(\pi e^2)(f \lambda)]$.  In the absence of
unresolved saturation, the apparent column density is a valid,
instrumentally blurred representation of the true column density
distribution.  Those regions of the profile exhibiting unresolved
saturated structure will be lower limits to the true column density.
The integration of the \nav\ profile is the total apparent column
density, which is reported in Table \ref{tab:totalcolumns}, along with
the equivalent widths of the absorption.  All of the upper limits
given here are $3\sigma$ following Wakker et al. (1996).  The \ovi\
absorption deserves some discussion, which we postpone to \S
\ref{sec:ovi}.  We have given values for \ciii\ and \oiv\ derived from
profile fits to the data for these ions (see below).  The \nav\
integrations, which are not as reliable as the profile fits for these
ions, give $\log N(\mbox{\ciii}) > 13.31$ and $\log N(\mbox{\oiv}) >
14.38$.


\begin{table}
\caption{Integrated Equivalent Widths and Column
  Densities$^a$ \label{tab:totalcolumns}}
\begin{tabular}{lrrccc}
\hline 
Ion & IP [eV]$^b$ & $\lambda_0$ [\AA] & $f$$^c$ & $W_r$ [m\AA] & $\log N_a$ \\
\hline
\HI   & 0 -- 13.6 &1215.670  & 0.4162 & $540\pm80$ & $14.07\pm0.06$$^d$ \\
\HI   & 0 -- 13.6 &1025.722  & 0.0791 & $90\pm19$ & $14.22\pm0.09$$^d$ \\
\HI   & 0 -- 13.6 & 972.537  & 0.0290 & $<50$ & $<14.31$$^c$ \\
\cii   & 11.3 -- 24.4 & 903.962  & 0.336  & $<32$ & $<13.12$ \\
\ciii  & 24.4 -- 47.9 & 977.020  & 0.757  & $89\pm9$ & $13.39\pm0.05$$^e$  \\
\nii   & 14.5 -- 29.6 &1083.990  & 0.111  & $<54$ & $<13.68$ \\
\niii  & 29.6 -- 47.4 & 989.799  & 0.123  & $<38$ & $<13.55$ \\
\niv   & 47.4 -- 77.5 & 765.148  & 0.616  & $29\pm9$ & $13.01^{+0.11}_{-0.16}$ \\
\oii   & 13.6 -- 35.1 & 834.465  & 0.132  & $<40$ & $<13.68$ \\
\oiii  & 35.1 -- 54.9 & 832.927  & 0.107  & $<36$ & $<13.73$ \\
%
\oiv   & 54.9 -- 77.4 & 787.711  & 0.111  & $106\pm8$ & $\ga14.37$$^f$  \\
\ov    & 77.4 -- 113.9& 629.730  & 0.515  & $218\pm8$ & $>14.39$  \\
\ovi   & 113.9 -- 138.1& 1031.926 & 0.1325 & $215\pm9$ & $14.47\pm0.02$$^g$  \\
\ovi   & 113.9 -- 138.1& 1037.617 & 0.0658 & $95\pm13$ & $14.50\pm0.05$$^g$  \\
\neviii  & 207.3 -- 239.1 & 770.409  & 0.1030 & $<29$ & $<13.73$ \\
\neviii  & 207.3 -- 239.1 & 780.324  & 0.0505 & $<25$ & $<13.97$ \\
\mgx   & 367.5 -- 367.5 & 624.950  & 0.041  & $<44$ & $<14.49$ \\
\svi  & 72.6 -- 88.1 & 933.378  &  0.437 & $<23$ & $<12.83$ \\
\hline
\end{tabular}
\medskip \\ {\em Note:} $a$: All upper limits are $3\sigma$.  Ions with rest
  wavelengths $\lambda_0 > 880$ \AA\ are from STIS observations, with
  $\lambda_0 < 880$ from \fuse\ observations.  
$b$: The range of creation and ionization energies for each species (D\"{a}ppen 2000).
$c$: Atomic oscillator strengths from Morton (2003) for
  $\lambda_0 > 912$ \AA\ and Verner et al. (1994) for $\lambda_0 <
  912$ \AA.  
$d$: We adopt a total \HI\ column density of $\log
  N(\mbox{\HI})=14.29\pm0.10$ based on a combined profile fitting
  analysis of the three transitions reported in this table.  The
  integrations here cover the entire breadth of the profile.  
$e$: Adopted from a component-fitting analysis.
$f$: Column density lower limit from a straight integration of the apparent 
  column density.  A component-fitting analysis gives $14.50\pm0.08$, but it 
  is sensitive to the assumed properties of the FUSE LSF.  See text.
  $g$: We adopt a total \ovi\ column density of $\log
  N(\mbox{\ovi})=14.47\pm0.02$ based on a weighted mean of the two
  lines, including absorption from the discrepant areas. The
  amount of \ovi\ associated with components 1, 2, and 3, is $\log
  N(\mbox{\ovi})=14.41\pm0.05$ (see \S \ref{sec:ovi}).  This is the value 
  used in our modeling.  
\end{table}

\begin{figure}
\includegraphics[width = 8truecm]{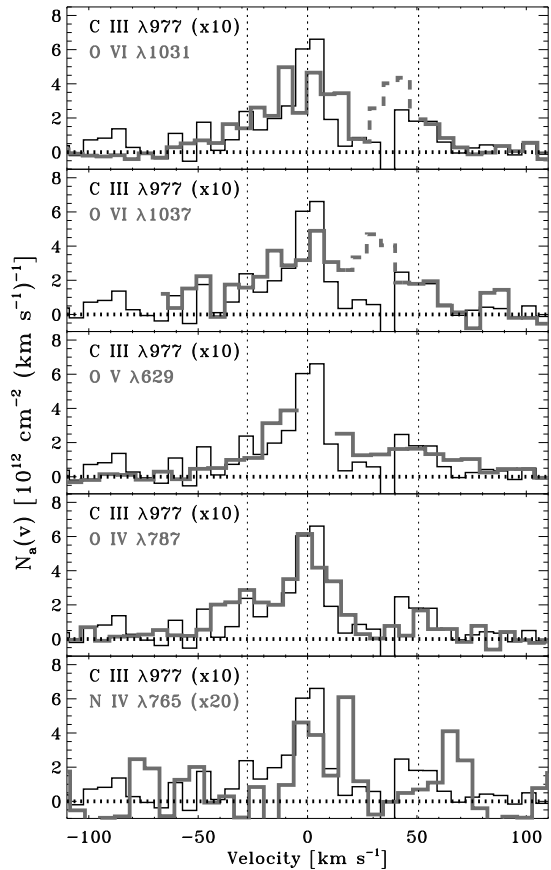}
\caption{Apparent column density profiles of metal ions in the
  $z=0.495096$ absorber toward \pks.  Both \ovi\ lines are shown given
  the differences between them (see \S \ref{sec:ovi}), which is shown
  by the dashed regions.  The profile for \ciii\ is shown in each
  panel, multiplied by a factor of 10 to match the scale of the other
  ions.  There is good agreement in the component structure between
  most of the ions.  The thin dashed vertical lines show the locations
  of the components fit to the \ciii\ profile (see \S
  \ref{sec:componentfits} and Figure \ref{fig:componentfits}).  The
  STIS observations (transitions with $\lambda_0>790$ \AA) are binned
  by a factor of two to approximately one datum per resolution element
  ($\sim7$ \kms).  The FUSE data have approximately the same pixel
  spacing, with three pixels per 20 \kms\ resolution element.
  \label{fig:nav}}
\end{figure}

We show the \nav\ profiles of several ions in Figure~\ref{fig:nav}.
With Figure~\ref{fig:metalprofiles}, these figures show the gas in
this system can be broken into three main absorption blends centred
at $v \sim -30, 0$, and $+50$ \kms, which we refer to as components 1,
2, and 3, respectively.  These components are most prominently viewed
in the profiles for \ciii, \oiv, and \ov.

\subsubsection{Profile Fitting of Metal Lines}
\label{sec:componentfits}

The \oiv\ and \ciii\ profiles show quite strong absorption in the
central component 2 ($v\sim0$ \kms), with peak apparent optical depths
of $\tau_a \approx1.4$ and 1.7, respectively.  Such large apparent
optical depths suggest there may be some unresolved saturated
structure in the profiles, making the measured apparent column
densities lower limits to the true columns.  An alternate way to
derive the column densities is to fit an instrumentally-smoothed model
to the absorption profiles of these species.  We use the profile
fitting software of Fitzpatrick \& Spitzer (1997) for this purpose.
The parameters for the fit are the central velocity, Doppler parameter
($b$-value), and column density of each assumed component or cloud.
The approach assumes each component can be described by a Maxwellian
velocity distribution.  

The fits depend on the adoption of an instrumental line spread
function (LSF).  For the STIS observations of \ciii\ $\lambda$977, we
adopt the STIS E140M LSF from Kim Quijano et al. (2003).  For the FUSE
observations of \oiv\ $\lambda$787 we adopt a Gaussian LSF with a FWHM
of 20 \kms.  The true breadth and shape of the FUSE LSF are uncertain
(see below).

\begin{figure}
\includegraphics[width = 8truecm]{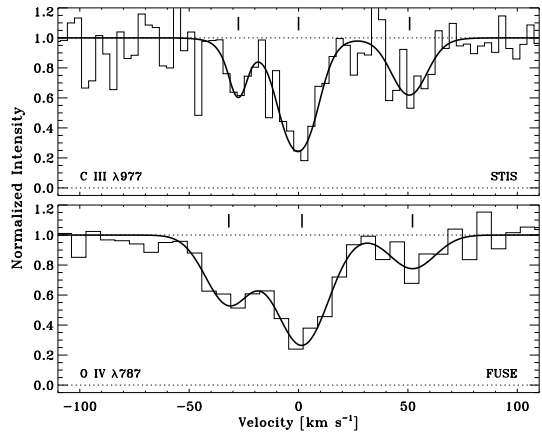}
\caption{The normalized absorption profiles of \ciii\ $\lambda$977 and
  \oiv\ $\lambda$787 with the adopted profile fits overlayed, assuming
  a redshift of $z=0.495096$.  The properties of the three components
  used in these fits are summarized in Table \ref{tab:profilefits}.
  The velocity centroids are shown by thick black ticks above each
  spectrum.  See the text for details of the fitting procedure.}
\label{fig:componentfits}
\end{figure}

The \ciii\ and \oiv\ profiles were both well fit by a three component
model, mirroring our discussion of the component structure above, with
reduced $\chi^2$ values very near unity.  The precise $\chi^2$ values
are sensitive to the range over which they are calculated, mostly
driven by the presence of large noise excursions (e.g., near
$v\sim-47$ \kms\ in the \ciii\ profile; see Figure
\ref{fig:metalprofiles}).  The results of the profile fitting for
these transitions, which we fit independently, are given in Table
\ref{tab:profilefits}.  Comparing the total columns derived in this
way with the \nav\ integrations (see previous subsection) suggests
significant corrections to the apparent columns due to saturation
effects are required.  Figure \ref{fig:componentfits} shows the
component models for each fit with the original data.


\begin{table}
\caption{Component Fitting Results\label{tab:profilefits}}
\begin{tabular}{lcccc}
\hline 
Ion & Component & $v_c$ [km/s]$^a$ & $\log N$ & $b$ [km/s] \\
\hline
\ciii 
& 1  & $-26.7\pm1.8$ & $12.56\pm0.15$ &  \nodata$^d$ \\
& 2  & $0.0\pm1.0$$^b$  & $13.20\pm0.06$ & $8.5\pm1.5$ \\
& 3  & $+50.7\pm2.1$ & $12.70\pm0.10$ & $8.9\pm3.8$ \\
\oiv  
& 1  & $-33.3\pm2.1$ & $13.90\pm0.08$ & $7.8\pm5.6$ \\
& 2  & $+1.5\pm1.3$  & $14.31\pm0.11$$^c$ & $9.1\pm2.9$ \\
& 3  & $+51\pm3$     & $13.47\pm0.11$ & \nodata$^d$ \\
\hline
\end{tabular}
\medskip \\ {\em Note:} $a$: Central velocity relative to $z = 0.495096$.
$b$: Our adopted redshift  is defined by the
  central velocity of this component.  
$c$: Given the low resolution of \fuse\ and the behavior
  of the profile fitting for this component, the quoted column density
  should be considered a lower limit.  
$d$: The $b$-value of this component is held fixed in the
  fitting to avoid fitting an unreasonably-low value.  The absorption
  is weak enough that the column density is not very sensitive to the
  adopted $b$-value.  
\end{table}


The \oiv\ results are derived from a fit of the \fuse\ data.  This
line is observed both by \fuse\ and STIS, although the STIS data have
very low S/N (see, e.g., Williger et al. 2006).  We have tested the
effects of simultaneously fitting the \fuse\ and STIS data and find no
differences in the results.  This is not surprising: because the fits
are weighted by the variance of the data, the low S/N of the STIS data
insures they do not contribute to the fits nearly as much as the
\fuse\ data.  However, the column density of \oiv\ can be sensitive to
the adopted width of the \fuse\ LSF, whose shape is
poorly-constrained, due to the change in the fitted $b$-value.
Furthermore, additional components in the \oiv\ profile can cause
significant increases in the column density and are not well
constrained on the whole.  A $\pm10$\%\ change in the FWHM of the
assumed LSF leads to changes in the total \oiv\ column of $-0.06$ and
$+0.10$ dex.  As the LSF breadth is increased, the errors in the
determination become quite large due to the increasing dependence on
the $b$-value with increasing saturation of the model.  (For an LSF
with FWHM$\, \approx 22$ \kms, the uncertainties are nearly an order
of magnitude.)  Given these uncertainties, we will proceed by adopting
the lower limit to the \oiv\ column from the \nav\ integration
discussed above.

The STIS observations of \ciii\ have significantly higher resolution
and a well understood LSF.  The contribution from extra components is
much more stringent for the \ciii\ profile, as they do not affect the
column density without increasing the $\chi^2$ value significantly
and/or requiring unphysical $b$-values (e.g., $\approx1-2$ \kms,
implying temperatures too low for the IGM) for the additional
components.  We feel that the column densities and errors derived from
our profile fitting analysis of the \ciii\ profile are good measures
of the true column.  In what follows we adopt the profile fitting
results for \ciii, as noted in Table \ref{tab:totalcolumns}.

\subsubsection{The O\,{\sevensize\it VI} Absorption}
\label{sec:ovi}

\begin{figure}
\includegraphics[width = 8truecm]{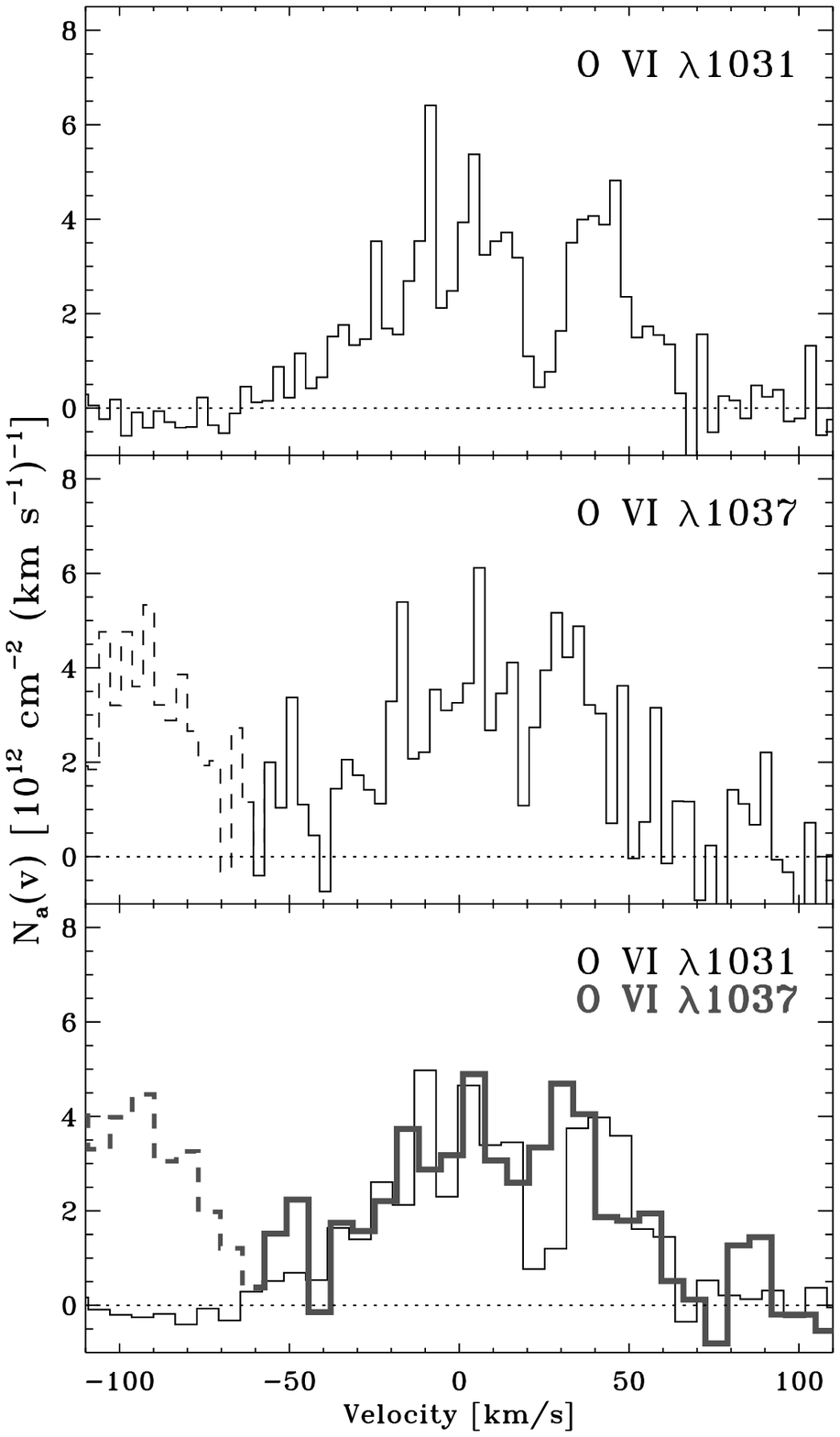}
\caption{Apparent column density profiles of the \ovi\ transition from
  the $z=0.495096$ absorber toward \pks.  The top two panels show the
  strong and weak lines, at 1031.926 and 1037.617 \AA, respectively,
  at full sampling.  The bottom panel shows both \nav\ profiles binned
  by two pixels (to one pixel per resolution element).  At velocities
  $15 \la v \la 45$ \kms, the discrepancies in the two profiles are
  caused by intervening transitions, presumably \lya. }
\label{fig:ovi}
\end{figure}

The \nav\ profiles of the weak and strong transitions of \ovi\ at
1031.926 \AA\ and 1037.617 \AA, respectively, are shown in
Figure~\ref{fig:ovi}.  For velocities $v \la +20$ \kms\ these
transitions are in good agreement.  However, they do not agree with
one another or the other metal ions at larger positive velocities (see
Figure \ref{fig:nav}).  A quick glance at Figures
\ref{fig:metalprofiles} and \ref{fig:ovi} reveals what appears to be a
return to the continuum in the strong line that is not present in the
weak line.  Examining the \nav\ profiles in Figure \ref{fig:ovi}, the
peak of column density distribution at velocities $v > +20$ \kms\
occurs at different velocities in the two profiles.  The ``component''
seen at $v\approx+30$ to $+40$ \kms\ in Figures \ref{fig:nav} and
\ref{fig:ovi} is not observed in other ions save perhaps \ov.
However, one does see component 3, at $v\approx+50$ \kms\ as seen in
other metal ion profiles, in both profiles.  This can be seen in
Figure \ref{fig:nav} (comparing, for example, the two \ovi\ profiles
with \oiv\ or \ciii).

The difference between the two members of the doublet could be due to
contamination of {\em both} profiles by intervening \lya\ absorption
with peak optical depths that yield consistant \nav\ integrations if
one assumes they are \ovi.  Limits on Ly$\beta$ absorption at the
appropriate redshifts do not rule this out.  Alternatively, as noted
by Tripp et al. (2008), this misalignment could be due to small scale
errors in the STIS geometric distortion correction, as have been
identified, e.g., by Jenkins \& Tripp (2001).  Such difficulties are
neither well characterized for STIS nor common.

While ``hot pixels'' can create discrepancies like those seen here
(see discussion in Tripp et al. 2008), the observations of \pks\ were
taken in two visits with the echelle orders displaced signficantly
between the visits.  The data extracted for the two visits
individually are consistent over these wavelength regions, which would
not be the case for hot pixels fixed on the detector.  Thus, this
explanation is not tenable.

We have attempted to simultaneously fit the profiles of the \ovi\
doublet as observed by STIS.  Aside from the velocity range of
possible contamination, the \nav\ profiles of the 1031 and 1037 \AA\
transitions seen in Figure \ref{fig:ovi} are in good agreement,
implying the effects of unresolved saturation cannot be large (as also
evidenced by the agreement in integrated apparent columns, see Table
\ref{tab:totalcolumns}).  The fitting was undertaken to attempt to
disentangle the component structure of the absorption and to provide a
measure of the column density of the higher velocity components,
especially component 3 ($v\approx+50$ \kms), which is strongly blended
with the contaminated or compromised regions.  We attempted fits under
two assumptions: 1) the discrepant regions in the \ovi\ lines are
contaminated by interloping absorbers; and 2) the discrepant regions
suffer from difficulties in the wavelength assignment, but represent \ovi\
absorption associated with this absorber.

\begin{figure}
\includegraphics[width = 8truecm]{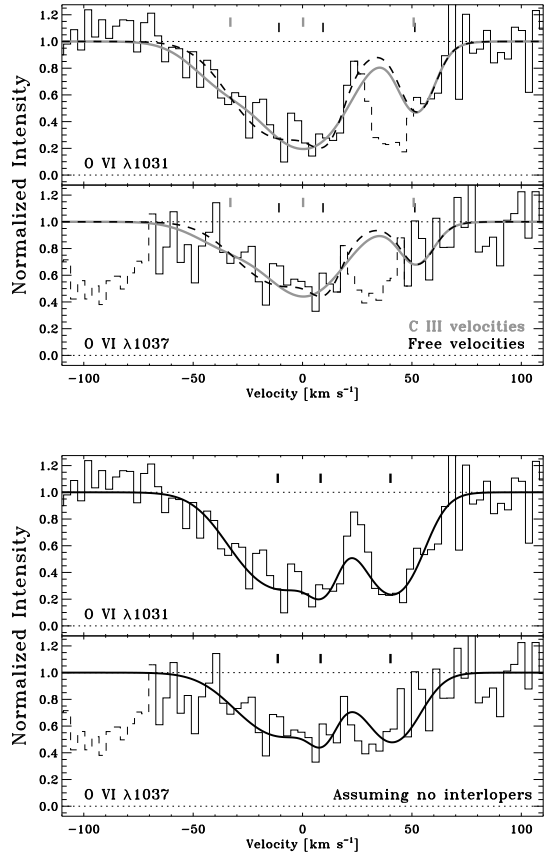}
\caption{The normalized absorption profiles of \ovi\
  $\lambda\lambda$1031 and 1037 with profile fits overlayed.  The top
  two panels show the profiles assuming the \ovi\ lines are
  contaminated with interloping absorbers (thin dashed lines denote
  regions of the data we assume to be contaminated). Two fits are
  shown in these panels: one assuming a three component fit with
  velocities fixed to those found in \ciii\ and the other allowing the
  velocities of the \ovi\ components to be a free parameter.  The
  reduced $\chi^2$ values are indistinguishable for these two models.
  The bottom two panels show our fits assuming there is no
  contamination of the profiles.  The $\chi^2$ is significantly worse
  in this case.  (However, the contribution to the $\chi^2$ values for
  the range of data assumed to be contaminated in the upper panels is
  minimized by artificially increasing the variance of the data over
  those ranges.)  The total column density for the models in the upper
  panels are both consistent with $\log N(\mbox{\ovi}) = 14.40$
  (albeit with much different errors; see text) whereas the bottom
  panels give $\log N(\mbox{\ovi}) = 14.50\pm0.13$.  The discrepancy
  is due to the lower equivalent width attributed to \ovi\ in the
  upper panels.}
\label{fig:ovicomponents}
\end{figure}

In the first of these fits, we artificially increased the variance
over the velocity ranges $+23 \le v \le +53$ \kms\ and $+19 \le v \le
+43$ \kms\ for the 1031 and 1037 \AA\ transitions, respectively, to
such large values that they do not affect the calculation of $\chi^2$
and, hence, the parameters of the fit.  The fits to the regions
encompassed by components 1 and 2 are non-unique, and can be fit with
one, two, or three components with similar values of $\chi^2$.  Part
of the difficulty in fitting these lower-velocity components is the
loss of information on the positive velocity side of the profiles due
to the assumed contaminating absorption.  In addition, the component
structure in the \ovi\ profiles is not distinct.  This is in contrast
to the \ciii\ and \oiv\ profiles in this region.  If the component
structure is assumed to follow that of \ciii, with velocities fixed to
those in Table \ref{tab:profilefits}, the total column density for
this model is $\log N(\mbox{\ovi}) = 14.41\pm0.05$.  The fit is shown
in Figure \ref{fig:ovicomponents} as the thick grey line, with the
central velocities of the components shown above the profiles.  A fit
in which velocities of the \ovi\ components to be a free parameter is
also shown as the thick dashed black line (the black ticks above the
profiles show the central velocities in this case).  The total column
density is $\log N(\mbox{\ovi}) = 14.40\pm0.16$.  In this fit, the
largest column density is in component 1; the large fitted $b$-value
for this component has a significant uncertainty, which is the source
of the much larger error in this model than that with the
velocities fixed to the \ciii\ solution.  Both fits have
indistinguishable values of $\chi^2$.  In each case the columns of
component 3 are very nearly the same near $\log N(\mbox{\ovi})_3 =
13.63\pm0.14$ (the value if the velocity is allowed to vary).

The second fit, which assumes all of the absorption is \ovi\
associated with this system, is shown in the bottom two panels of
Figure \ref{fig:ovicomponents}.  These plots show the nature of the
difficulty quite well for the positive velocity component.  Components
1 and 2 are virtually indistinguishable from the fits above where the
velocities are allowed to vary.  The total column density in the \ovi\
fit is $\log N(\mbox{\ovi}) = 14.50\pm0.13$, some $+0.1$ dex higher
than that derived above.  We somewhat favor this total column over the
lower value.  The total equivalent width for the full profile is
$W_r(\lambda1031\ \mbox{\AA}) = 215\pm9$ m\AA\ (see Table
\ref{tab:totalcolumns}), which places this absorber in the top 10\% of
the $\approx50$ intervening \ovi\ absorbers studied in the recent
survey of Tripp et al. (2008).  The column and equivalent width values
in this assumption are consistent with those derived by Tripp et
al. (2008) and Thom \& Chen (2008b).

We focus our ionization analysis on the gas associated with components
1, 2, and 3, excluding the additional column over the discrepant
region in \ovi\ that does not appear in \ciii\ and \oiv.  Because the
discrepant component is not seen in \ciii\ or \oiv, it must have
significantly different ionization conditions than the rest of the
gas, and we do not attempt to model it.  Because most of our models
will assume a solar C/O ratio, metallicities [C/H] derived from \ciii\
will also imply a given [O/H].  Thus for our ionization and
metallicity calculations (when they depend on \ovi) will adopt the fit
fixed to the \ciii\ velocities, which gives $\log N(\mbox{\ovi}) =
14.41\pm0.05$.  This excludes a component at $v\approx+40$ \kms, but
includes absorption from component 3 at $v\approx+50$ \kms\ (top
panels in Figure \ref{fig:ovicomponents}).

This value is preferred over that in which the velocities of the \ovi\
are allowed to freely float because the uncertainties become
unreasonably large in the latter case in our estimation.  An
integration of the \nav\ profile of \ovi\ over the uncontaminated
region at $v\la+20$ \kms\ yields $\log N(\mbox{\ovi}) = 14.30\pm0.05$
(a mean of the two \ovi\ lines).  Adding the column associated with
component 3 from above, this gives a total column density of $\log
N(\mbox{\ovi}) = 14.38\pm0.05$ excluding the compromised regions
(i.e., for components 1 to 3).  Thus, the uncertainty derived using a
model with velocities fixed to those found in \ciii\ provides an
uncertainty closer to the \nav -derived values.  We note that
including absorption associated with component 3 only impacts the
metallicities -- the ratios of \ovi /\ciii\ and \oiv /\ciii\ derived
for components 1 and 2 are consistent with the ratio after including
component 3, which adds $\sim+0.1$ dex to these three ions.  Thus, had
we chosen to model the clearly-uncompromised velocity range, we would
derive the same results as those presented below.

\subsubsection{Limits on the Multiphase Structure from Velocity Profiles}
\label{sec:multiphasekinematics}

The \nav\ profiles shown in Figure \ref{fig:nav} demonstrate that the
ions \ciii, \oiv, \ov, and \ovi\ in this absorber have a similar
velocity structure, with gas associated with the three components
discussed above (and perhaps a fourth seen in \ovi\ and \ov).  The
profile fitting results of \S \ref{sec:componentfits} reinforce this
basic structure, though with the caveat that the limited resolution of
the data could be hiding substructure within the components (or
component blends) identified in Table \ref{tab:profilefits}.  The \HI\
and metal ions would be considered ``well-aligned'' in the
categorization of Tripp et al. (2008) since their central velocities
are consistent with one another within $2\sigma$.  However, the
uncertainties in the \HI\ centroid are relatively large, so the
meaning of this alignment is not clear.  In addition, too little
information is available for Lyman-series absorption to know if the
detailed component structure hinted at in the metal ions is mirrored
in \HI.  

While the metal line profiles in Figure \ref{fig:nav} are generally
similar, there are hints of differences between them.  The broad
envelope of the \ovi\ distribution (at $v \la +20$ \kms) is similar to
the \oiv\ and \ciii\ distributions, and the total velocity extents of
\ovi, \oiv, and \ciii\ are all very similar (including the gas in
component 3 at $v \approx +50$ \kms).  However, \ovi\ seems less
prominent over the velocity range of component 2 compared with the
lower-ionization species.  This may be due to changing ionization
conditions between components 1 and 2, with a corresponding change in
the ionic ratios, or due to somewhat broader components in \ovi.
These differences, however, are at the limits of the noise and
resolution of the current data.  It is difficult to say whether or not
they are significant.  If so, they may imply the existence of a warmer
phase of the gas traced by \ovi\ and a cooler phase traced by \ciii\
and \oiv.  However, if this is true, {\em both phases must be present
  at the same velocities}.

The component Doppler parameters given in Table \ref{tab:profilefits}
limit the temperature of the gas.  For example, the $b$-values of
\ciii\ and \oiv\ for component 2 imply maximum temperatures of $T \la
(5.2\pm1.8)\times10^4$ K and $ \la (8\pm5)\times10^4$ K, respectively.
\footnote{The profile fits \oiv\ are uncertain as discussed above.
  Lowering the FWHM of the assumed LSF for FUSE to 18 \kms\ allows for
  temperatures $\la(1.0\pm0.5)\times10^6$ K.}  The contribution of a
hot component to these species must not produce absorption profiles
broader than observed.  To test whether a significant amount of
warm-hot gas (with $T>10^5$ K) could be included in these profiles, we
have fit the \ciii\ and \oiv\ profiles including an additional
component forced to have a $b$-value appropriate for gas with
$T\ga10^5$ K.  For \ciii, we find that any broad components added in
this way must have negligible column densities in order to be
consistent with the observed profiles.  This is not unexpected given
the small ionization fraction of \ciii\ in gas with such temperatures
(see below).  For \oiv, one can include a broader component centred
near $\approx0$ \kms.  Adopting a $b$-value of 10 \kms, corresponding
to $T\la10^5$ K for such a component leads to very small column
densities in that component or forces the existing component 2 to have
unphysically-low $b$-values (implying temperatures much lower than
expected for the IGM) for reasonable values of $\chi^2$.  If one
assumes $b=16$ \kms, corresponding to $T\la2.5\times10^5$ K, the fits
are significantly better.  Our fits suggest at most 30\% of the total
column of \oiv\ can come from such a component.  This limit comes
about because as the column of the broad component is increased above
$\log N(\mbox{\oiv})\approx13.7$, the column of the existing component
2 increases as well (due to a decreasing $b$-value for component 2).
This conclusion is robust to changes in the adopted LSF for FUSE.  It
is mostly driven by the nature of a multicomponent curve of growth,
where the $b$-value of the narrow component is decreased in order to
account for an increased contribution to the equivalent width from the
broad component.  The decrease in the $b$-value for the narrow
component increases the total column density substantially, always
keeping the broad component to $\la30\%$ of the total.

Thus, given the profile comparisons and the profile fitting results,
there may be room, if not some evidence, for a multiphase structure
within this absorber.  The discrepancies between the \nav\ profiles of
the ions that might imply a multiphase structure are subtle compared
with those seen in many absorbers (e.g., Tripp et al. 2008), and the
\ovi\ and other ions are reasonably well aligned with the \HI\ in this
system (see Figure \ref{fig:metalprofiles}).  If broad, warm-hot
($T\ga10^5$ K) gas is present, it can only contribute $\la30$\% of the
total column of \oiv\ and must have a negligible contribution to the
\ciii\ column density.  In such a scenario, \ovi\ may preferentially
trace the warm-hot material, \ciii\ would trace the cooler gas (at
$T\la5\times10^4$ K), and \oiv\ would include contributions from both,
albeit with a larger contribution from the cooler matter.  The
warm-hot gas traced by \ovi\ in such a scenario cannot be too hot ($T
\la 3\times10^5$ K) in order to avoid producing too much \neviii\ or
too little \ov\ (see \S \ref{sec:ionization}).

\subsection{H\,{\sevensize\bf I} Lyman-series Absorption}
\label{sec:hydrogen}

\begin{figure}
\includegraphics[width = 8truecm]{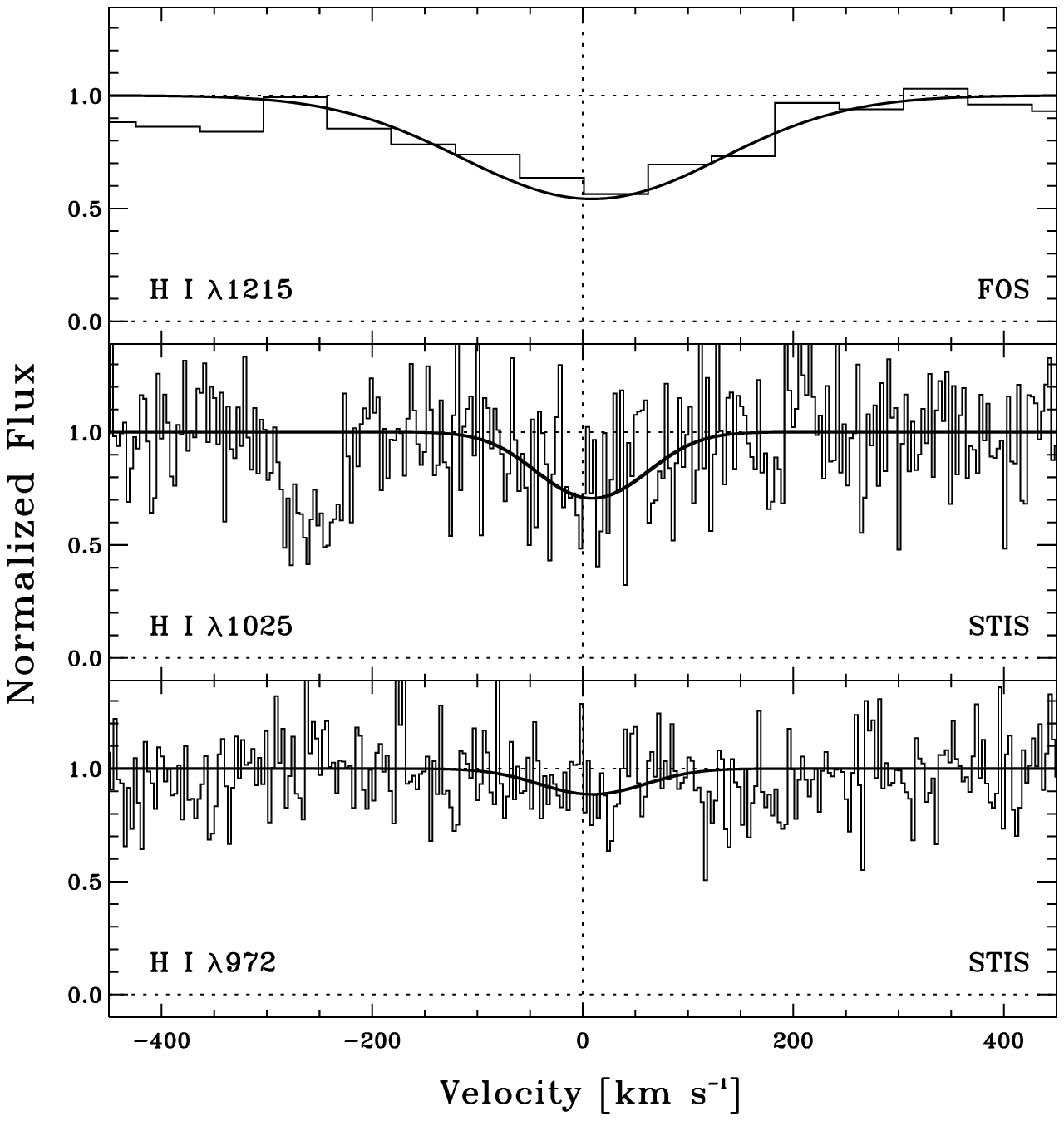}
\caption{Absorption profiles of the \HI\ \lya, \lyb, and \lyg\
  transitions associated with the $z=0.495096$ absorber toward \pks\
  (the velocity zero-point was defined by the peak of the \ciii\
  absorption).  The \lya\ absorption is from the low-resolution FOS
  spectrum, while the other two are from the high resolution STIS
  data. The histograms show the data, while the smooth solid lines
  show a simultaneous profile fit to the three lines (see \S
  \ref{sec:hydrogen}).}
\label{fig:hydrogen}
\end{figure}

Figure \ref{fig:hydrogen} shows the normalized absorption line
profiles from the \lya, \lyb, and \lyg\ transitions of neutral
hydrogen for the $z=0.495$ absorber.  The \lya\ observations are taken
from the low-resolution FOS data ($\approx230$ \kms\ FWHM), while the
\lyb\ and \lyg\ are from the high-resolution STIS data ($\approx7$
\kms\ FWHM).  The \lya\ transition is poorly resolved by the FOS, the
\lyb\ line is poorly detected by STIS, and \lyg\ is undetected.  We,
therefore, have very little information on the velocity structure of
the \HI\ in this system.

Table \ref{tab:totalcolumns} reports our integrated restframe
equivalent widths and limits for these transitions.  The \lya\ line
integration is over the range $\pm350$ \kms, while the integration for
\lyb\ and \lyg\ is over $\pm85$ \kms.  To better constrain the \HI\
column in the face of the resolution and detection limitations, we
have performed a simultaneous profile fit to these three transitions
as described above (see also Appendix in Lehner et al. 2007 for this
absorber).  We adopt a Gaussian line spread function (LSF) for the FOS
data and allow for a shift in the FOS data relative to the STIS data
given the better absolute wavelength calibration of the latter.

The results of the fit are shown in Figure \ref{fig:hydrogen} as the
smooth solid lines.  We note that the results of the fit are somewhat
sensitive to the range over which one calculates the $\chi^2$ goodness
of fit parameter.  We have investigated the range of column densities
implied as one varies these limits, incorporating an estimate of this
variability into our error estimate.  From this analysis, we adopt a
best-fit column density of $\log N(\mbox{\HI}) = 14.29\pm0.10$.  This
yields a $b$-value of $b=68\pm8$ \kms\ for a single component fit to
the profiles with a central velocity of $v_c = +9\pm8$ \kms\ relative
to $z=0.495096$ (from \ciii).  

Our adopted \HI\ column is comparable to the values derived from
earlier studies that include this absorber: $\log N(\mbox{\HI}) =
14.31\pm0.07$ (Tripp et al. 2008), $14.3\pm0.1$ (Thom \& Cen 2008b),
$14.39\pm0.07$ (Williger et al. 2006), and $> 13.95$ (Bahcall et
al. 1993, using their \lya\ equivalent width).  The central velocity
is driven by the \lyb\ detection shown in Figure \ref{fig:hydrogen};
obviously there is significant uncertainty in this determination given
the poor quality of that profile.  While the derived velocity is
within $\sim1\sigma$ of the central value of \ciii, \ovi\ and other
ions, the poor quality of the \HI\ determination limits our ability to
draw significant conclusions from this alignment.  As discussed above,
the \HI\ would be described as ``well-aligned'' under the Tripp et
al. (2008) definition since the central velocities of \HI\ and \ovi\
are within $2\sigma$ of one another.  The $b$-value provides a firm
limit to the weighted mean temperature of the \HI -bearing gas of
$T\la (2.8\pm0.7)\times10^5$ K assuming pure thermal broadening.  The
$b$-value assumes a single component describes the \HI\ broadening,
though our data are not of high enough quality to determine if the
\HI\ component structure follows that of the metals (with multiple
components).

\section{Ionization of the Absorber}
\label{sec:ionization}

Given the importance of \ovi\ as a tracer of shock-heated gas in the
low-redshift IGM, the presence of strong \ovi\ in the $z=0.495$
absorber toward \pks\ might imply this system is tracing the WHIM at
low-redshift.  However, as emphasized by a number of authors (Tripp et
al. 2008, Thom \& Chen 2008b, Lehner et al. 2006, Prochaska et
al. 2004, Savage et al. 2002), \ovi\ may not be a pure tracer of
shock-heated WHIM material.  If not associated with the WHIM at
$T>10^5$ K, it may trace cooler shock-heated gas, the metal-enriched
\lya\ forest, or photoionized regions in the extended halos of
galaxies.

There are two principal mechanisms for the ionization of the gas being
considered here: photoionization by the UV background and collisional
ionization.  The characteristics of photoionized gas depend on the
shape and strength of the ionizing spectrum and on the density of the
gas, while the properties of collisionally ionized gas depend
principally on the temperature of the gas.  Gas that is predominantly
photoionized will have characteristic temperatures of order a few
times $10^4$ K.  Collisional ionization models of IGM absorbers are
usually constructed to explain highly ionized gas and adopt
temperatures $\sim10^5$ K and above (e.g., Gnat \& Sternberg 2007, Cen
et al. 2006), although lower temperatures are affected by collisions
as well.

The presence of strong \ovi\ in the absorber being studied here may
arise due to the photoionization of a low density absorber, to the
presence of hot gas or gas that has cooled from high temperatures in a
non-equilibrium manner, or from a mixture of these.  In this section
we consider pure photoionization and pure collisional ionization
models for the gas making up this absorber, showing that neither type
of model fits the data sufficiently well.  We also investigate an
admixture of photoionized and collisionally ionized material, a
scenario with some support from the comparison of the \ovi\ and \ciii\
or \oiv\ velocity structure.  

Though there are likely regimes in which either photoionization or
collisional dominates, IGM gas will have contributions from both of
these mechanisms: photons will be present in intergalactic space and
collisional ionization is important for at least hydrogen above
$\sim10^4$ K.  Sophisticated models that account for the impact of
both are beyond the scope of this paper (see Wiersma, Schaye, \& Smith
2008, Cen \& Fang 2006 for recent attempts at such models within large
scale simulations).  We consider a small number of models in which
high temperature gas is exposed to ionizing photons, though ours is
not an extensive exploration of that approach.

We assume that the observed \HI\ and metal ions in this absorber are
nearly cospatial and that the absorber has uniform abundances.  We use
solar relative abundances as a starting point for comparing ions of C,
N, O, Ne, and Mg to models.  Unfortunately, there is some controversy
about the relative abundances of these elements in the solar system.
Meteoritic abundances of Mg are well-determined and give a good
abundance measure for the solar system (with the Mg to Si abundance in
meteorites being referenced to the photospheric Si abundance to
provide an absolute Mg abundance); we adopt $\log {\rm Mg/H}=-4.46$
(Lodders et al. 2009).  The abundance of Ne is typically measured
relative to O in the corona, and we assume Ne/O$\, = 0.15$ or $\log
{\rm Ne/O} = -0.82$ (e.g., Asplund, Grevesse, \& Sauval 2005a, Basu \&
Antia 2008).

Solar system abundance estimates for C, N, and O depend on
photospheric abundances (see recent overviews by Lodders et al. 2009,
Asplund et al. 2005a, and Lodders 2003).  Recent adoption of 3D
hydrodynamic models in analyzing the photospheric data have led to
significantly lower abundances (see Asplund 2005) than earlier results
(e.g., Anders \& Grevesse 1989).  This is particularly true for O, for
which Asplund et al. (2004) derive $\log {\rm O/H} = -3.34\pm0.05$
compared with the earlier standard of $\log {\rm O/H} = -3.17\pm0.06$
(e.g., Grevesse \& Sauval 1998).  Such a low abundance causes
significant difficulties with helioseismology models (e.g., Bahcall et
al. 2005, Basu \& Antia 2008), and there are arguments in favor of the
higher abundances (e.g., Ayres 2008, Delahaye \& Pinsonneault 2006,
and others).

We will proceed by following the recent critical summary of abundances
by Lodders et al. (2009) and assume $\log {\rm O/H} = -3.27\pm0.07$
(an average of values from Caffau et al. 2008, Ludwig \& Steffen 2008,
and Melendez \& Asplund 2008).  With our adopted solar system Ne/O
ratio, this implies $\log {\rm Ne/H} = -4.09$.

The abundance of C advocated by Lodders et al., $\log {\rm C/H} =
-3.61\pm0.04$ (an average of results from Allende-Prieto et al. 2002,
Asplund et al. 2005b, and Scott et al. 2006) is less controversial.
Lodders et al. recommend $\log {\rm N/H} = -4.14\pm0.12$ from Caffau
et al. (2009); though this result has not yet appeared in press we
adopt this value for consistency.  For comparison, Asplund et
al. (2005a) advocate $\log {\rm N/H} = -4.22\pm0.06$.  For Mg we adopt
$\log {\rm Mg/H}=-4.46\pm0.06$ (Lodders et al. 2009).


\subsection{Collisional Ionization Models}

Ionization of metals by collisions (primarily with electrons) can lead
to high-stage metal ions if the gas temperature is sufficiently high.
This is what drives the WHIM models, in which the high temperatures
are a result of high velocity shocks as gas accretes onto filaments,
groups, and clusters of galaxies (e.g., Cen \& Ostriker 1999, Dav\'{e}
et al. 1999, Furlanetto et al. 2004, Kang et al. 2005, Bertone et
al. 2008).  These works emphasize the potential importance of \ovi\
for tracing this material, though several cosmological simulations
also predict that some of the \ovi\ likely traces cool, photoionized
gas (e.g., Cen et al. 2001, Fang \& Bryan 2001, Chen et al. 2003, Kang
et al. 2005, Oppenheimer \& \dave\ 2009).

Given the strong \ovi\ in the absorber studied here, it is worth
considering if the ionization of the absorber can be explained
primarily through collisional ionization.  We use the recent models of
Gnat \& Sternberg (2007) to test whether collisional ionization can
explain the relative distribution of ionization states in the
$z=0.495$ absorber toward \pks.  These authors have calculated new
collisional ionization calculations with up-to-date atomic
physics. They discuss models under the assumption of collisional
ionization equilibrium (CIE) as well as non-equilibrium (NEQ) cooling
models in which gas initially at $T=5\times10^6$ K cools isochorically
or isobarically.

\begin{figure}
\includegraphics[width = 9truecm]{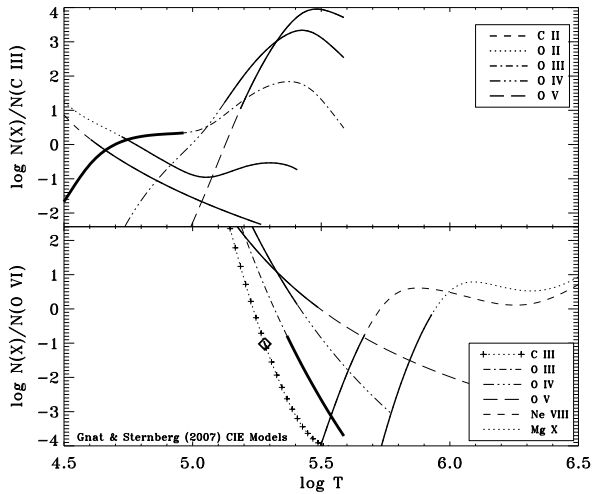}
\caption{Column density ratios of several ions relative to \ciii\
  ({\em top}) and \ovi\ ({\em bottom}) as a function of temperature
  for the CIE models of Gnat \& Sternberg (2007).  The solid lines
  represent regions for which each of the ratios is consistent with
  the observations of the $z\approx 0.495$ absorber toward \pks.  Note
  that the \oiii /\ciii\ and \oiii /\ovi\ ratios cannot be
  simultaneously matched for any temperatures.  Furthermore, the \ciii
  /\ovi\ ratio is not matched for temperatures consistent with the
  line width of the \ciii.  (The narrow range of temperatures for
  which \ciii /\ovi\ is matched is denoted by the diamond in the
  bottom panel.)  The full range of temperatures is not shown in each
  panel because of the limited range of temperatures over which Gnat
  \& Sternberg find significant amounts of \ciii\ and \ovi.  The NEQ
  ratios have similar distributions at [M/H]$\, = -1$.  Those at solar
  abundance have significantly more \ovi\ at lower temperatures and
  similar difficulties matching all of the available ionization
  states.  }
\label{fig:gnatcie}
\end{figure}

One may argue against single-temperature CIE models for this absorber
without detailed calculations, as the profile fits to the \ciii\
profiles suggest temperatures well below the $T\ga10^5$ K needed to
produce significant \ovi\ absorption in CIE.  A comparison of the
ionic ratios for the species in Table \ref{tab:totalcolumns} shows no
single temperature for which all of the data can be matched by the CIE
models.  This is demonstrated in Figure \ref{fig:gnatcie}, which shows
the column density ratios predicted by the Gnat \& Sternberg CIE
models for a number of species, identifying temperatures over which
the models are consistent with the data for each of the ratios.  The
difficulties are principally in simultaneously matching the ratios of
\ciii\ to \ovi\ and limits on \oiii\ to \ciii.  Prochaska et
al. (2004) and others have emphasized that \ciii\ and \ovi\ cannot
coexist in significant amounts in single-temperature CIE models.  The
temperature limits derived for the \ciii\ in this absorber provide
support for this conclusion.

The column densities may be better matched if one assumes a
two-temperature structure for the absorber, with both phases in CIE.
In this ad hoc model, one assumes that the \ovi\ traces a warm-hot
phase and that the \ciii\ traces a separate warm phase, perhaps a
low-temperature WHIM like that discussed by Kang et al. (2005). The
upper and lower panels of Figure \ref{fig:gnatcie} show the
constraints for each of the independent temperature regimes.  The
temperature of the \ovi -bearing medium is constrained to be in the
range $2.3\times10^5 \la T \la 4.7\times10^5$ K by the upper limits
for \oiii\ and \neviii\ at the low and high end of the range,
respectively.  In this case the \ciii -bearing gas, if in CIE, would
itself require $T\la10^5$ K based on the \oiii\ to \ciii\ ratio,
consistent with the \ciii\ $b$-values.  In this scenario, \oiv\ and
\ov\ are present in both temperature regimes, though most of these
ions must reside in the \ovi -bearing phase for the \ciii -bearing
phase to be at $T\la10^5$ K.  This is very difficult to reconcile with
the breadth of the \oiv\ profile, and the limit derived in \S
\ref{sec:multiphasekinematics} for the fraction of that profile that
can be associated with hot gas.  Of course one may posit a
distribution of temperatures, but the models will become increasingly
complex and difficult to justify in the face of the relatively simple
assumptions that one makes when adopting CIE models.  In view of these
arguments, pure CIE models are not a good representation of the
ionization of this absorber.

Shock-heated WHIM gas may not, however, be in equilibrium.  If hot gas
cools faster than it can effectively recombine, the gas may be
over-ionized for its temperature (e.g., Shapiro \& Moore 1976, Edgar
\& Chevalier 1986, Sutherland \& Dopita 1993, Heckman et al. 2002); in
addition, if the ion and electron temperatures of shock-heated gas are
significantly different, NEQ conditions also hold (see Cen \& Fang
2006, Yoshikawa \& Sasaki 2006, Bertrone et al. 2008).  The
comparison of the relative timescales for cooling, ionization, and
recombination is a measure of whether such NEQ ionization is
important.  Tripp et al. (2008) discuss these timescales for the IGM;
unfortunately, they could not conclusively argue for or against NEQ
effects being important for the WHIM.  Recently, Wiersma et al. (2008)
have argued that photoionization of metal coolants can suppress
cooling in the IGM, increasing the cooling time significantly.  If so,
this argues NEQ cooling effects are likely not important.  

Gnat \& Sternberg (2007) have produced NEQ models that follow the
ionization fractions of metal ions as gas cools from an initial
temperature of $T=5\times \e{6}$ K.  The resulting ionization
fractions are metallicity dependent.  We have considered models with
solar and 10\% solar metallicity (see \S \ref{sec:photoionization} and
\S \ref{sec:multiphase}).  For many species, the NEQ models approach
the CIE models at low metallicity (Gnat \& Sternberg 2007; Tripp et
al. 2008), \ovi\ being one of those for which NEQ models at [M/H]$\, =
-1$ are quite similar to CIE models.  Given this similarity, it is not
surprising that the NEQ models have nearly the same difficulties as
the CIE models.  The \ciii /\ovi\ ratio is only matched at
$T\approx1.9\times \e{5}$ K for the Gnat \& Sternberg
isobaric\footnote{Gnat \& Sternberg 2007 discuss the conditions over
  which which isobaric or isochoric models are likely to be
  appropriate.  This absorber is near the cross-over point, though
  more likely to be in the isobaric regime.  Isochoric models give
  very similar results.} models at both metallicities, inconsistent
with the \bvalues\ for \ciii\ and with the \oiii\ upper limits.
Two-temperature NEQ models can be constructed as discussed above for
the CIE models, but all require most of the \oiv\ to be associated
with gas at temperatures $\sim2\times \e{5}$ K, which is inconsistent
with our component fitting analysis.  For solar abundance models, the
limit to the \oiii /\ciii\ ratio is requires $T<2\times10^4$ K in the
NEQ models, inconsistent with other constraints in this case (e.g.,
from \oii).


Thus, it is unlikely that the gas making up the $z=0.495$ absorber
toward \pks\ is dominated by collisional ionization.  This holds when
considering either CIE or NEQ models.  If our observations had only
measured \HI, \ciii, and \ovi, a common situation in IGM observations,
these models could possibly be questioned by considering the component
structure for this absorber.  However, given the low-S/N ratio of
these and other STIS observations of the low-\z\ IGM, it would be
difficult to strongly rule them out.

\subsection{Photoionization Models}
\label{sec:photoionization}

The $z=0.495$ absorber is exposed to the ultraviolet background (UVB)
from the integrated light of QSOs and galaxies in the universe.  As
such, photoionization will play a role in determining the ionization
of the gas, which is optically thin to ionizing radiation, even if
collisions are the dominant ionization mechanism (Wiersma et
al. 2008).  Here we consider models in which collisional ionization is
unimportant, and the ionization state of the absorber is determined by
photoionization alone.  In this scenario, the absorber amounts to a
metal-enriched \lya\ forest cloud.

For optically-thin systems, the ionization state of photoionized gas
is primarily determined by the ratio of the ionizing photon to
hydrogen volume densities, i.e., the ionization parameter $U \equiv
n_\gamma/n_H$, and the spectral shape of the ionizing background.  The
metallicity can have a minor effect on the ionization through its
importance for the thermal balance.  We use the Cloudy ionization code
(version 07.02; last described by Ferland et al. 1998) to model the
photoionization of a plane-parallel slab of gas illuminated by a
diffuse UVB.  Because this is intrinsically a one dimensional model,
it is effectively an infinite thin sheet diffusely illuminated on both
sides by the UVB.  We stop the calculations when the integrated \HI\
column matches that observed.  Our models vary the density for the
assumed ionizing spectra (see below), which is akin to varying the
ionization parameter.  We assume solar relative abundances for the
initial models with a base metallicity of [O/H]$\, = -0.5$ and adjust
the metallicity at a later point to best match the total column
densities of the ions.  The metallicity plays only a small role in the
relative ratios of the metal ions, although it fixes the ratio of
metal ions to \HI.

We investigate two possible models for the ionizing background at
$z\approx0.5$ from the Haardt \& Madau (2005, in preparation; details
given in Haardt \& Madau 2001) update to the work by Haardt \& Madau
(1996).  These background spectra, calculated with the CUBA software
(Haardt \& Madau 2001), assume that: 1) the UVB is dominated by
quasars and active galactic nuclei (the QSOs-only spectrum); or 2) the
UVB is a combination of light from quasars and integrated light
escaping from galaxies (the QSOs+galaxies spectrum).\footnote{One
  could imagine that the absorber were near to an individual galaxy
  and estimate the ionizing spectrum seen from that galaxy as a
  function of distance as, for example, Fox et al. (2005) have done
  for studying high velocity clouds about the Milky Way.  The present
  models attempt to study the effects of the integrated light escaping
  from the overall population of galaxies.} The input QSO spectrum in these
models uses a power law index of $\alpha=1.8$ for wavelengths below
1050 \AA\ (e.g., Zheng et al. 1998, Telfer et al. 2002), while the
basic spectral shape for the light leaking from galaxies is based on a
spectrum from Bruzual \& Charlot (1993, 2003) libraries for a
population of stars with 0.2 solar metallicity and 0.5 Gyr age.  The
ionizing ($h\nu>1$ Ryd) radiation coming from the galaxies is
attenuated by the gas and dust within the galaxies.  The escape
fraction of ionizing gas from the galaxies is an important quantity
for calculating the background, and Haardt \& Madau adopt $f_{\rm
  esc}=0.1$ in their base model (F. Haardt, 2008, priv. comm.).  As
noted by Haardt \& Madau (2001), the definition of the escape fraction
in this model is not the ratio of escaping Lyman continuum photons to
those produced by the underlying stellar population.  Rather, this
value is normalized to the observed (i.e., dust attenuated) 1500 \AA\
flux of galaxies (see Eqn. 1 in Haardt \& Madau 2001).  In the
QSOs+galaxies spectrum, the majority of the hydrogen ionizations are
caused by photons that have escaped from galaxies.

\begin{figure*}
\includegraphics[width = 18truecm]{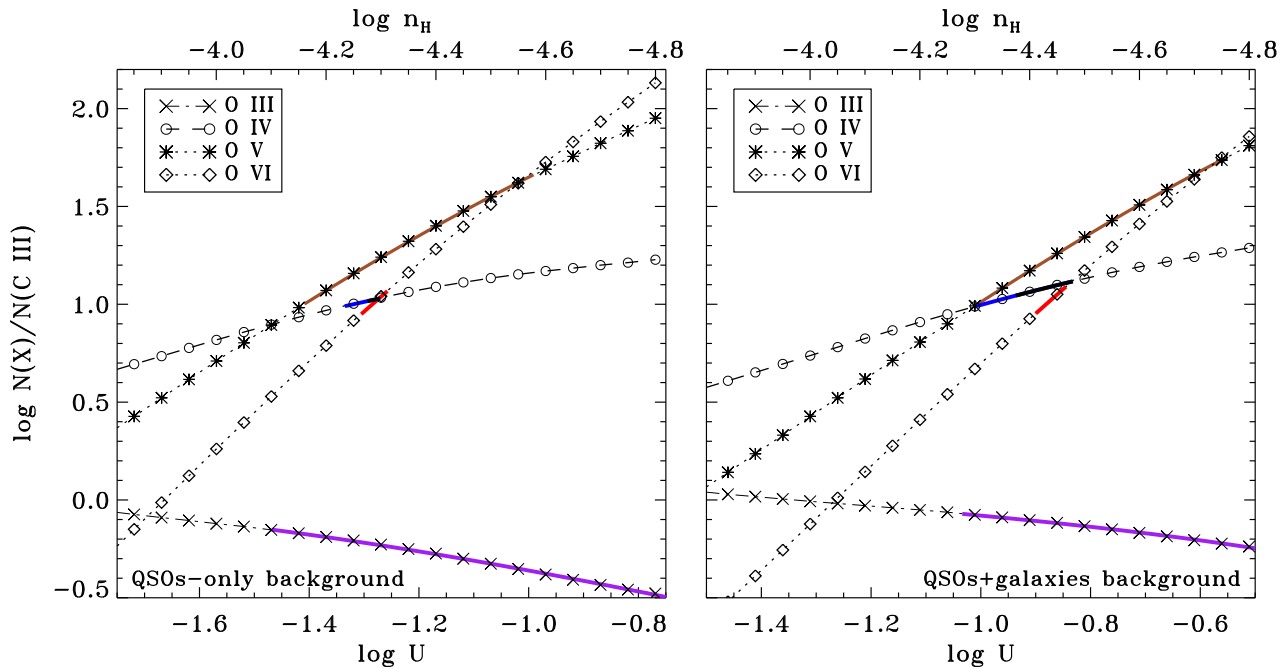}
\caption{Photoionization models for the $z=0.495$ absorber toward
  \pks\ assuming an ionizing background dominated by QSO radiation
  ({\em left}) or a background that also contains a contribution from
  galaxies ({\em right}).  The column density ratios of several ions
  relative to \ciii\ are plotted, with the solid lines representing
  regions over which the models are consistent with all of the
  observed column density ratios and limits (i.e., not just with
  \ciii\ as plotted here).  For O IV, we show the region allowed if
  one adopts the component fitting results as the thick grey line.
  For our assumed UVB intensities, the densities implied by the
  ionization parameters are given across the top of each plot.  The
  models shown assume a metallicity of [M/H]$=-0.15$ and $-0.62$ for
  the QSOs-only and QSOs+galaxies backgrounds, respectively.  The
  metallicity has a small effect on the ionization balance and, hence,
  only a small impact on these plots which show relative metal ion
  column densities.  We assume solar relative abundances for C/O.  The
  best-fit models assuming a QSOs-only background give $\log U\approx
  -1.28$ while the models with a QSOs+galaxies background give $\log
  U\approx-0.87$.}
\label{fig:cloudymodels}
\end{figure*}

Figure \ref{fig:cloudymodels} shows the results of our photoionization
models for the QSOs-only ({\em left}) and QSOs+galaxies ({\em right})
spectra.  We plot the ratio of column densities of \oiii, \oiv, \ov,
and \ovi\ to \ciii\ from the Cloudy models as a function of the
ionization parameter assuming solar relative abundances.  The thick
solid lines denote regions of the models for which the calculated
ratios are consistent with observed column densities.  While we have
plotted the $N(X)/N(\mbox{\ciii})$ ratios, the thick lines are for the
individual ions are based on all of the available constraints.  These
plots do not include all of the ions for which we have limits, as many
of the model values are well below the upper limits for this absorber.
Also, the ions of nitrogen are not used to constrain the ionization.
Non-solar N/C or N/O ratios are common at moderately-low
metallicities, and we use the \niv\ column to estimate the N/O ratio
in the absorber.  When using \oiv\ we have been conservative, since we
are concerned that the component fitting results may not fully account
for the saturation in the profile; we adopt the lower limit derived
from integrating the \nav\ profile of \oiv.  The models that fall
within $1\sigma$ of the component fitting column are also shown with a
black line in Figure \ref{fig:cloudymodels}. 

The best-fit models for the QSOs-only spectrum has $\log
U\approx-1.28\pm0.03$, which matches all of the column density
constraints from Table \ref{tab:totalcolumns}.  The quoted
uncertainties in the ionization parameter represent the range over
which the models are within $1\sigma$ of the observed ratio.  The
constraints on the models rely on most of the available ions, but the
most important are the \ovi /\ciii, \oiii /\oiv, and \oiv /\ovi\
ratios.  The constraints are significantly less stringent without the
inclusion of the \oiii\ limits.  

The metallicity implied by these models is $[{\rm O/H}] = [{\rm C/H}]
= -0.15\pm0.07$, where the uncertainty accounts for the range of
ionization parameters and observational uncertainties, but does not
attempt to account for any systematic uncertainties associated with
the choice of UVB, adopted solar system abundances, missing atomic
data, or model assumptions.  For our adopted UVB, these models have
$n_{\rm H}\approx5.0\times10^{-5}$ \percc\ and $\log N({\rm
  H})\approx18.5$, giving a pathlength for the model cloud of $\sim20$
kpc.  The derived density is similar to the predictions for average
\lya\ forest cloud of this column density from the models of Schaye
(2001) and \dave\ et al. (1999), both of which predict $n_{\rm
  H}\approx10^{-5}$ \percc.  The simulations show a large scatter in
the $\rho / N(\mbox{\HI})$ relation and have factor of a few type
uncertainties.  The neutral fraction of hydrogen for this model is
$x(\mbox{\HI})\equiv N(\mbox{\HI})/N({\rm H}) \approx6\times10^{-5}$.
Cloudy predicts equilibrium temperatures $T\approx28,000$ K for both
assumed UVB models.

The best-fit models based on the QSOs+galaxies spectrum match the
available constraints for $\log U\approx-0.87\pm0.03$, corresponding
to $n_{\rm H}\approx3.5\times10^{-5}$ \percc\ for this spectrum.  The
metallicity implied by this model is $[{\rm O/H}] = [{\rm C/H}] =
-0.62\pm0.07$, nearly 0.5 dex lower than that derived using the
QSOs-only UVB.  This model has $\log N({\rm H})\approx19.0$, giving a
pathlength for the model cloud of $\sim90$ kpc.  Thus, the fraction of
neutral hydrogen is $x(\mbox{\HI})\approx2\times10^{-5}$.  The larger
ionization fraction of H and ionization parameter in the QSOs+galaxies
models comes about because the radiation leaked from galaxies
contributes significantly more photons between 1 and 4 Rydbergs.
Thus, to produce \oiv\ and \ovi\ columns comparable to those in the
QSOs-only models requires more H-ionizing photons (and \ciii -ionizing
photons).

For both choices of UVB, comparing the measured \niv\ column density
with those predicted by the models gives $[{\rm
  N/O}]\approx-0.60^{+0.11}_{-0.16}$.  This value of [N/O] is
consistent with [N/O] seen in metal-poor galaxies and damped \lya\
systems at similar metallicities (see Henry \& Prochaska 2007, Pettini
et al. 2008). It is difficult to assess the significance of this
result given the uncertainties in both the models and the data.
Though, given the significant differences in total metallicity between
the two UVB models, it is encouraging that both give the same [N/O]
results.

We have assumed a solar C/O ratio in the above analysis.  Danforth et
al. (2006) have made some arguments for ${\rm [C/O] }\, \approx -1$ in
low-\z\ \ovi +\ciii\ systems (their equation 15; see also Danforth \&
Shull 2008), although they view this result with a great deal of
skepticism.  We have enough information from the oxygen ions alone to
test whether the present absorber has such a low value.  Using only
the oxygen ions to constrain models using the QSOs-only UVB, the
ionization parameter is constrained to be $\log U = -1.36\pm0.10$.
This model gives ${\rm [O/H] } = 0.0\pm0.2$, i.e., solar abundance,
with ${\rm [C/O] } = -0.2\pm0.1$ and ${\rm [N/O] } = -0.7\pm0.2$.
Adopting the QSOs+galaxies model leads to $\log U = -0.92\pm0.10$.
This model gives ${\rm [O/H] } = -0.5\pm0.2$, ${\rm [C/O] } =
-0.1\pm0.1$ and ${\rm [N/O] } = -0.7\pm0.2$.  Thus, for these assumed
UVB, the data do not support a C/O ratio as low as 0.1 times solar.

A slightly sub-solar C/O abundance is consistent with our models,
although solar C/O is not ruled out.  Allowing the C/O to be subsolar
affects the derived [O/H] abundances moderately.  Notably, the
QSOs-only model produces a solar oxygen abundance, which seems
unlikely for an absorber arising in the diffuse IGM; however this
absorber, given its large \ovi\ column, may arise near to a galaxy.
We will discuss the implied abundances further in \S
\ref{sec:discussion}.

\subsection{Multiphase Models}
\label{sec:multiphase}

We have shown above that a pure photoionization model can match the
constraints provided by the ionic column densities in this absorber.
Even the \ovi\ can be adequately explained by photoionization by the
diffuse UVB.  The \ciii\ and \oiv\ profiles are well fit with
relatively narrow components (see Table \ref{tab:profilefits})
implying low temperature gas consistent with photoionization, although
we have some concern about the limits from the \oiv\ fits as discussed
above.  The component structure of \ovi\ is not as clearly delineated.
This may simply be due to the low signal-to-noise of the observations,
but could also imply broader components in \ovi\ than the lower ions.
The \ovi /\ciii\ ratio varies somewhat with velocity in the profile,
albeit at moderate significance.  Furthermore, there is room within
the \oiv\ profile for a broad component tracing warm-hot gas
($T\ga10^5$ K) that contributes as much as $\sim30\%$ of the \oiv\
column.  The breadth of the \ciii\ profile suggests high-temperature
collisionally-ionized gas cannot be important for that ion (and,
indeed, this is not expected; see Prochaska et al. 2004).  While the
evidence is not firm, this absorber could plausibly contain regions of
differing metal ion ratios and perhaps different ionization mechanisms
and temperatures (\S \ref{sec:multiphasekinematics}).  The evidence
for multiphase absorption is certainly not as clear-cut as some cases,
for example, emphasized in the Tripp et al. (2008) survey of low-\z\
systems.

Here we consider a multiphase model for the ionization of the
$z\approx 0.495$ absorber.  In our model, the column densities are
predicted using a linear combination of the pure collisional
ionization and photoionization models considered above (see, e.g.,
Prochaska et al. 2004).  The free parameters are the ionization
parameter of the photoionized component, the temperature of the
collisionally ionized component, the fraction of the total hydrogen
column associated with the collisionally ionized medium, \fraccoll{H},
and the metallicity of the gas, assumed to be the same between phases.
This is meant as a crude model of a cool, photoionized cloud
surrounded by shock-heated hot gas.  We note that this model is not
self-consistent in that neither is the collisionally-ionized phase is
itself not subjected to an ionizing radiation field nor is the
photoionized phase subject to radiation from the hot,
collisionally-ionized phase.  We are thus assuming that collisional
processes so dominate the ionization of the hot phase that the UVB
does not have a significant impact on this phase while the radiation
from that hot phase is not intense enough to rival the UVB in
importance for the photoionized phase.

\begin{figure*}
\includegraphics[width = 15truecm]{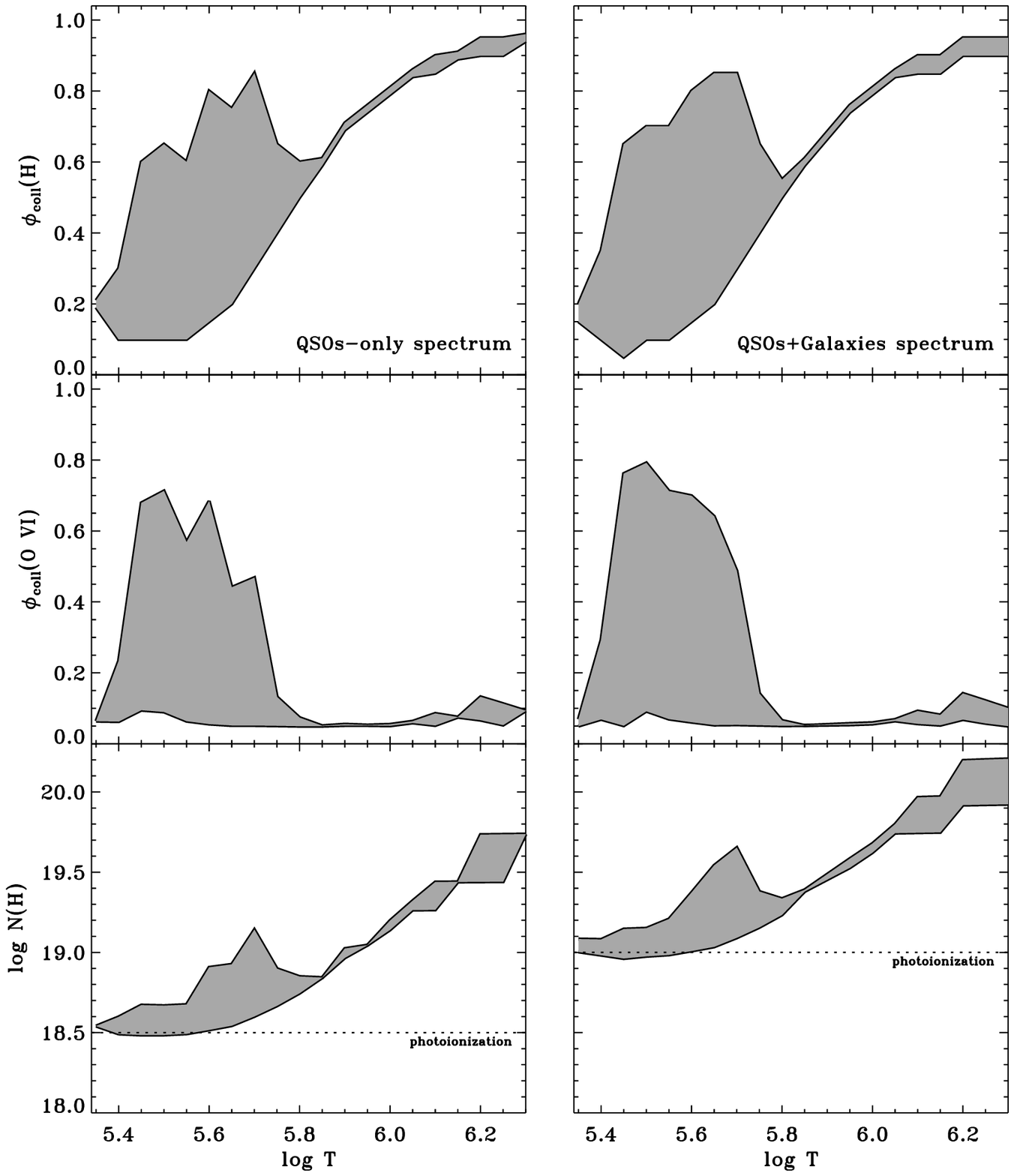}
\caption{Allowed regions of parameter space for multiphase models
  assuming \fraccoll{\oiv}$\, \le 0.3$.  For both the QSOs-only UVB
  ({\em left}) and QSOs+galaxies UVB ({\em right}) we show the range
  of allowed values for \fraccoll{H}, \fraccoll{\ovi}, and the total
  $\log N({\rm H})$.  The values of $\log N({\rm H})$ at $\log T \ga
  5.7$ are largely set by the limits on \neviii\ and \mgx\ under the
  assumption of collisional ionization.  At these temperatures, these
  are the only ions that provide constraints on the collisionally
  ionized phase, although we have required \fraccoll{\ovi}$ \,\ge
  0.05$.  The \fraccoll{H} and $N({\rm H})$ curves are upper limits
  for temperatures $\log T\ga5.7$ if one does not require the gas to
  contribute significantly to the \ovi\ profiles.  These figures
  include H columns from both the photoionized and
  collisionally-ionized phases.  The columns predicted by a pure
  photoionization model are shown as the dotted lines.  The total
  column density limits for the WHIM are summarized below.}
\label{fig:multiphase}
\end{figure*}

The results of these models are shown in Figure \ref{fig:multiphase},
which shows the range of allowed values for \fraccoll{H},
\fraccoll{\ovi}, and $\log N({\rm H})$ as a function of temperature
for models adopting the QSOs-only or QSOs+galaxies spectra of Haardt
\& Madau (2005, in preparation) and Gnat \& Sternberg (2007) CIE
models.  Here \fraccoll{H} and \fraccoll{\ovi} are the fractions of
the total H and \ovi\ columns that arise in the hot,
collisionally-ionized gas.  These models assume that $\la30\%$ of the
\oiv arises in the collisionally-ionized phase (i.e.,
\fraccoll{\oiv}$\, \le 0.3$). If this limit is relaxed, a broader
range of contributions from the hot, collisionally-ionized phase is
allowed for temperatures $\log T \la 5.6$, although the higher
temperature results are not much affected.  These models assume CIE,
but the NEQ models give results that are not significantly different.
Our calculations extend over the temperature range $4.5 \le \log T \le
6.5$, although models for $\log T \la 5.30$ ($T\la2\times10^5$ K) did
not produce models in agreement with the observations.  The ionization
parameters are not shown because they differ little from the values
derived in the pure photoionization models discussed above (\S
\ref{sec:photoionization}).  Because we are interested in the origins
of the \ovi, these plots only show models for which
\fraccoll{\ovi}$\,\ge0.05$.  This is the cause of the apparent lower
limit to the total H column densities.  If we do not require that the
collisionally-ionized phase contribute to the \ovi\ column densities,
the lower right of the top and bottom panels in these figures would
both be allowed (see \S \ref{sec:discussion}).

We find multiphase models that are consistent with our observational
constraints for $T\ga2.2\times10^5$ K ($\log T \ga 5.35$) for both
adopted UVB spectra.  The ionization parameters appropriate for the
photoionized phase are similar to those reported in \S
\ref{sec:photoionization}.  They differ from the pure photoionized
models by up to $-0.2$ dex and $+0.05$ dex, i.e., the maximum
excursions are to lower ionization parameter.  For temperatures
$T\sim2.8\times10^5$ K to $5\times10^5$ K ($\log T\approx5.45$ to
5.70) we have very little constraint on the fractions of H or \ovi\
arising in the collisionally-ionized gas, largely because we only have
limits to the \oiv /\ovi\ ratio.  Thus, there is a degeneracy between
the collisionally-ionized phase and the photoionized phase for the
oxygen ions where one phase may compensate for the other.  The \ciii\
still arises almost exclusively in the photoionized phase.

The metallicities in these mixed-phase models are very similar to
those derived from pure photoionization models.  The range of allowed
metallicities for the QSOs-only UVB models is [O/H]$\, \sim -0.33$ to
$-0.05$; for the QSOs+galaxies UVB we find [O/H]$\, \sim -0.76$ to
$-0.52$.  This is quite insensitive to the assumptions for
$T\ga5\times10^5$ K ($\log T\ga5.7$) since the collisionally-ionized
phase contributes little to the detected ions, which are all
associated with the photoionized phase save for a small fraction of
\ovi.

Thus, a simple multiphase model with distinct collisionally-ionized
and photoionized regions allows for significant amounts of hot gas.
For temperatures $T\ga5\times10^5$ K this is due to the nature of the
observational constraints.  We do not detect \neviii\ or \mgx, and
these are the only ions covered by our observations with significant
ionization fractions at these temperatures.  At temperatures
$T\sim2.8\times10^5$ K to $5\times10^5$ K ($5.45 \la \log T \la
5.70$), the fraction of the total column of hydrogen allowed to be
associated with the collisionally ionized phase may be significant and
is poorly constrained.  We find none of these models match all of the
available constraints for a collisionally-ionized phase at
temperatures $T\la2\times10^5$ K.

Another type of multi-process model is one in which gas at a fixed
temperature is irradiated by the UVB (e.g., Tripp et al. 2008,
Danforth et al. 2006).  In this approach, we fix the gas temperature
within our Cloudy models, which provides for collisional ionization,
and calculate the ionization balance for a range of ionization
parameters at each temperature.  This is effectively a
collisionally-ionized gas that is modified by the UVB ionization.  For
both UVB models investigated here the models are only able to match
the available constraints for $T\la 40,000$ K, consistent with the
$b$-values of components seen in \ciii.  The main difficulty is in
avoiding overproduction of \oiii\ at high temperatures.  The allowable
ionization parameters, set by the \ovi /\ciii\ ratio, are similar to
those discussed above, with $\log U \approx -0.85\pm0.05$.  This model
also gives a similar abundance with [O/H]$\, \approx -0.70\pm0.05$.
Similarly, the models assuming a QSOs-only UVB spectrum give
[O/H]$\,\approx -0.27\pm0.05$ at $\log U \approx -1.23\pm0.05$.  A
single temperature collisionally-ionized gas exposed to the UVB, then,
is inconsistent with our observations for the large temperatures
consistent with the WHIM, $T\sim10^5$ to $10^7$ K.  Indeed, the
temperatures allowed by this approach are generally so low that they
approach those produced by a pure photoionization model (which gives
$T\sim28,000$ K).

\section{Discussion}
\label{sec:discussion}

We have presented measurements of five metal ions and limits on
another eight in the $z\approx0.495$ absorber toward \pks.  We have
discussed a range of possible ionization mechanisms, determining that
this absorber may be described by a pure photoionization model, but
may also harbor some collisionally ionized material at $T \ga 10^5$ K
if it is mixed with a cool photoionized component.

\subsection{The WHIM and Low-\z\ O\,{\sevensize\bf VI} Systems}
\label{sec:discussionsub1}

Intervening \ovi\ absorption line systems are quite common at low
redshifts, with $dN/dz \approx 10-20$ for $W_r \ga 30$ m\AA\ (Danforth
\& Shull 2008; Tripp et al. 2008; Thom \& Chen 2008) depending on the
criteria used to select the \ovi\ systems.  Thus, they probe a
significant baryon reservoir no matter their origin.  However, they
have elicited the most interest due to the possibility that they trace
the ``missing baryons'' associated with a WHIM.  For systems with $W_r
\ga 200$ m\AA\ like the one studied here, the absorber density is
significantly lower, of course, with $dN/dz \approx 2$ (Tripp et
al. 2008).

Some intervening absorbers with significant \ovi\ absorption likely do
trace the WHIM.  These may be absorbers with signatures of pure
collisionally ionized gas at high temperatures, such as the systems at
$z\approx0.31978$ toward PG 1259+593 from Richter et al. (2004) or at
$z\approx0.1212$ toward H1821+643 discussed by Tripp et al. (2001), in
which \ovi\ and \HI\ have breadths consistent with
$T\approx(2-3)\times10^5$ K.  These may also be found in absorbers
that are part of a multiphase structure including a very hot component
($T\ga5\times10^5$ K) such as the \neviii -bearing system at
$z\approx0.207$ toward HE 0226-4110 (Savage et al. 2005).  There are
reasonable arguments for the origins of other absorbers in hot
shock-heated gas, such as the $z\approx 0.056$ systems toward PKS
2155--304 that Shull et al. (2003) have argued trace material
infalling onto a group of galaxies (which may exhibit X-ray
absorption; Fang et al. 2002).  However, for most \ovi\ absorbers, the
connection to the WHIM is less evident or non-existent (Prochaska et
al. 2004, Lehner et al. 2006).  In their survey of $z\la0.4$ \ovi\
systems, Tripp et al. (2008) find $\ga34\%$ of their intervening
absorbing components are associated with cool gas at $T<10^5$ K based
on an analysis of $b$-values.  Thus, a significant fraction are
obviously too cool to be associated with the $T>10^5$ K WHIM, although
they may represent cooled WHIM material (e.g., Kang et al. 2005).
However, $\sim50\%$ of that sample shows significant differences
between the \HI\ and \ovi\ profiles, suggesting a multiphase
structure.  Such absorbers could include gas at $T\ga10^5$ K.

While the origins of \ovi\ in the IGM may be ambiguous, with
photoionization being a significant contributor to the \ovi\ in a
significant number number absorbers (e.g., Tripp et al. 2008, Thom \&
Chen 2008b, Oppenheimer \& \dave\ 2009, Kang et al. 2005), the Li-like
ions \neviii\ and \mgx, which peak in abundance in CIE models at
$T\sim 6.4\times \e{5}$ K and $1.1\times\e{6}$ K, respectively (Gnat
\& Sternberg 2007), are much more secure probes of shock-heated hot
gas.  These ions are probes of WHIM material at temperatures where a
significant fraction of the low-\z\ baryons are expected to reside
(\dave\ et al. 2001) and probe gas in the temperature regime
accessible via X-ray absorption studies ($5\times10^5 < T <
3\times10^6$ K).  The absorber considered here is one of the few \ovi\
absorbers for which \neviii\ could be searched (Prochaska et al. 2004,
Richter et al. 2004, Savage et al. 2005, Lehner et al. 2006), and the
only one for which a limit on \mgx\ is currently available.  A summary
of previous observations of \neviii\ in the low-\z\ IGM is given in
Table 7 of Lehner et al. (2006).  \neviii\ has only been detected in
the multiphase absorber at $z\approx0.207$ toward HE~0226$-$4110
(Savage et al. 2005), the system with the second highest
$W_r(\mbox{\ovi} \ 1031) = 169\pm15$ m\AA\ among those yet searched
for \neviii\ (the highest being the current absorber).  Savage et
al. derive $N(\mbox{\neviii})/N(\mbox{\ovi}) = 0.33\pm0.10$ for this
absorber with $\log N(\mbox{\ovi}) = 14.37\pm0.03$.

Our $3\sigma$ limit to the \neviii\ column density gives the lowest
limit on this ratio yet observed, $N(\mbox{\neviii})/N(\mbox{\ovi}) <
0.18$ ($3\sigma$) assuming the \ovi\ derived from integrating the full
profiles (see \S \ref{sec:ovi}); a slightly higher value of $<0.21$ is
derived if using the \ovi\ in the velocity ranges of components 1, 2,
and 3.  The \ovi\ equivalent width of this system is 25\%\ larger than
that for the $z\approx0.207$ absorber toward HE~0226$-$4101.

\begin{figure*}
\includegraphics[width = 15truecm]{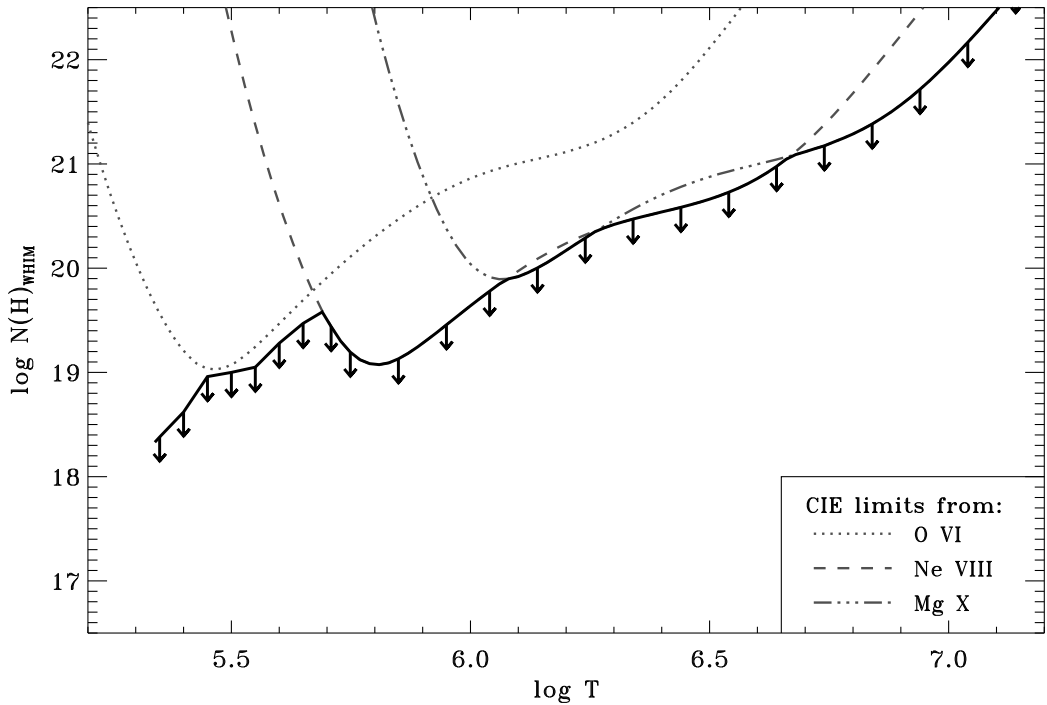}
\caption{Limits on the collisionally-ionized WHIM column density of
  this absorber as a function of temperature based on limits on the
  column densities of \ovi, \neviii, and \mgx\ assuming CIE ionization
  fractions with an abundance of [M/H]$\, = -0.62$ (see \S
  \ref{sec:photoionization} and \S \ref{sec:multiphase}).  The
  species-dependent limits for CIE are shown with the dashed curves.
  At temperatures $\log T < 5.7$, the limits are from the multiphase
  analysis discussed in \S \ref{sec:multiphase} (see also Figure
  \ref{fig:multiphase}).  These limits are dependent on the
  assumptions of those models.  We note that for $6.4 \la \log T \la
  7.0$, \ovii\ has an ionization fraction of nearly unity in the
  models considered.  Over that range the column of this ion is simply
  4.4 dex below the column of hydrogen plotted here for our assumed
  abundances.  Higher assumed abundances will lower the H column
  density limits.}
\label{fig:whimlimits}
\end{figure*}

The lack of \neviii\ and \mgx\ absorption in the $z\approx 0.495$
absorber toward \pks\ implies that there is not a very large reservoir
of hot [$\sim(0.5 - 3)\times\e{6}$ K], collisionally ionized gas
associated with this absorber, although the precise limits depend on
the temperature of the gas.  Such hot gas would not be detectable in
the other ions probed by our observations due to its high temperature
(and hence ionization state).  Figure \ref{fig:whimlimits} shows
limits to the total H column density as a function of temperature for
gas associated with the WHIM in this absorber assuming
[M/H]$\approx-0.62$ (from the models adopting a QSOs+galaxies UVB).
The distribution of these limits with temperature arises from jointly
considering the \ovi, \neviii, and \mgx\ ionization fractions and
observational limits on their columns.  The limits from the individual
ions assuming CIE are shown by the gray lines.  At $\log T < 5.7$, we
assume the results from Figure \ref{fig:multiphase} for a multiphase
absorber.  The H column density limits displayed in Figure
\ref{fig:whimlimits} are lower if one assumes a higher metallicity
(e.g., from models with the QSOs-only UVB).

Two things are important from this figure.  First, if additional hot
gas is present with temperatures $\log T\la6.5$, it would not provide
sufficient column to lead to detectable X-ray absorption except
perhaps along the sight lines to the very brightest sources.  Over the
range $6.0 \le \log T\le 6.5$, we infer the \ovii\ K$\alpha$ line
should have an equivalent width $W_r < 6$ m\AA\ to $W_r < 15$ m\AA\
with a column $\log N(\mbox{\ovii})\la 16.7$ (all $3\sigma$ limits).
These are derived from the limits on \neviii\ and \mgx, assuming
assuming CIE ionization fractions and solar relative abundances (the
absolute abundance is not important since the \ovii\ column is
predicted from other metal species).  We also assume pure thermal
broadening of the lines.  \oviii\ K$\alpha$ would be even weaker, with
a maximum of $W_r < 5$ m\AA\ and $\log N(\mbox{\ovii})\la 15.75$ for
$\log T \sim6.5$.  The presence of \ovii\ at the equivalent width
limit for the highest temperatures might be marginally detectable
toward bright objects (e.g., Williams, Mathur, \& Nicastro 2006, Fang
et al. 2007), although this is a $3\sigma$ limit and absorption is
very unlikely to be present at that level.  Furthermore, the
discrepancies discovered between various instruments and groups for
X-ray absorption lines reported at good significance toward extremely
bright objects makes it difficult to imagine the current absorber
could be reliably detected if at $\log T \la 6.5$ (e.g., see
discussion in Bregman 2007).  Second, the limits for WHIM material at
$\log T\ga6.5$ are quite limited due to low ionization fractions of
even these highly-ionized species at such temperatures.  The majority
of the metal-bearing WHIM is thought to be at temperatures below this
(e.g., \dave\ \& Oppenheimer 2007), with higher temperature IGM gas
tracing large overdensities associated with clusters.  At $\log T\sim
7$, both the \ovii\ and \oviii\ transitions are limited to $W_r \la
20$ to 25 m\AA\ ($3\sigma$).  We note that photoionization models for
this absorber, which can match the available constraints quite well,
predict $\log N(H)\sim18.5$ to 19.0.

The Li-like ions limit the column of very hot gas in this absorber
over a temperature range at which the majority of the WHIM baryons are
expected to reside to be relatively low.  And while some of the gas in
this absorber could be associated with a few$\times10^5$ K gas, a
significant cool photoionized component is needed to explain the lower
ionization species such as \ciii.  The \ovi\ in this system may be
described fully by a photoionized gas.  However, even if some of the
\ovi\ is produced via collisional ionization, the absorber must
include some photoionized material (with $\sim50\%$ or more of the
\ovi\ coming from a cool photoionized phase).  Because of the
uncertainties in its origins, \ovi\ is not a robust, pure tracer of
the WHIM in this absorber.  There is no direct evidence for hot gas
with $T\ga 10^5$ K material in this absorber, even though it has very
strong \ovi.  For temperatures $T\ga 5\times10^5$ K, however, the gas
would only be seen in \neviii\ or \mgx\ absorption.

The absorber studied here may have a multiphase structure, where the
high-temperature shock-heated gas contributes somewhat more than half
of the total H column density (unless at $T\ga5\times10^5$ K, in which
case we have few constraints on the gas).  Tripp et al. (2008) note
that $\approx50\%$ of their intervening absorbers show strong
differences between the \HI\ and \ovi\ profiles, evidence for a
possible multiphase structure.  The evidence from the velocity
profiles of the absorption for a multiphase structure to this system
is slight at best.  While the \ovi /\ciii\ seems to vary somewhat
between components 1 and 2, this does not necessarily imply these
components have significantly different ionization mechanisms and
temperatures.

The multiphase absorbers noted by Tripp et al. typically show large
discrepancies in the velocity distribution of \HI\ and \ovi\
absorption and/or differences in the characteristics of the detected
low-ionization metals compared with \ovi.  If the present absorber
does have a mixture of phases, it may indicate that the fraction of
the low-\z\ \ovi\ systems with multiphase structure is larger than
Tripp et al. estimate.  The presently available observations from STIS
and \fuse\ do not probe a broad enough range of ions or measure the
velocity-resolved profiles with enough signal-to-noise to constrain
the structure of many systems.  Future observations with the Cosmic
Origins Spectrograph will provide much higher quality spectra,
allowing us to detect weaker lines and study the velocity profiles at
higher signal-to-noise (albeit at lower resolution than STIS).  In
particular, observations of \civ\ would provide significant diagnostic
power (see below); better quality observations of the \HI\ transitions
could allow a search for broad components therein.

Simulations generally predict that the high \ovi\ equivalent width
systems like the present absorber are the most likely to arise in the
hotter gas, either the diffuse WHIM or gas in higher overdensity
regions including galaxy halos (Oppenheimer \& \dave\ 2009, Ganguly et
al. 2008, Kang et al. 2005, \dave\ et al. 2001).  The environment of
the absorber is pertinent to its origins, and a number of works have
explored the connection between \ovi\ absorbers and galaxies (e.g.,
Wakker \& Savage 2009, Cooksey et al. 2008, Lehner et al. 2008, Stocke
et al. 2006, Tripp et al. 2006, Tumlinson et al. 2005, Sembach et
al. 2004).  Prochaska et al. (2006) have presented a galaxy redshift
survey for the sight line to \pks\ (see also Williger et al. 2006).
The surveys of this sight line are so far inadequate to bear on the
relationship of this absorber to galaxies.  Prochaska et al. find only
three $L\sim6L_*$ galaxies within 1000 \kms\ of the absorbers.  Two of
these galaxies are found within $\sim200$ \kms\ of the absorber, but
at impact parameters $\rho \ga 10$ Mpc.  This survey is $>80\%$
complete to $R=21$ (or roughly $L \sim 0.8L_*$) and $70\%$ complete to
$R=22$ ($\sim0.3L_*$).  No galaxies aside from the three high
luminosity systems are found at this redshift.  So, if this system is
associated directly with a galaxy, the galaxy is likely to be of order
$L\sim0.1L_*$ or less.  This absorber is not at so high a redshift
that it is impractical to eventually study the fainter galaxies near
this absorber with 8-m to 10-m class telescopes.

\subsection{Abundances and Ionization Mechanisms in the Low-\z\ IGM}

The metallicity of low-\z\ IGM absorbers has been the subject of
much work over the last decade.  The abundance distribution of the IGM
can be used to probe the extents over which galaxies expel metals,
whether the metals escape the galaxies or fall back onto them, and the
impact of such metal expulsion on galaxy evolution (e.g., \dave\ \&
Oppenheimer 2007, Calura \& Matteucci 2006, Tumlinson \& Fang 2005).
The level of metal enrichment then suggests the sphere of influence
over which galaxies affect the IGM, and it is an important
consideration when using metal lines to calculate the baryon density
of low-\z\ IGM absorbers since in this case $\Omega_b \propto Z^{-1}$
(e.g., Tripp et al. 2008, Thom \& Chen 2008a, Danforth \& Shull 2008).

Many low-redshift metal line absorbers tend to show metallicites
between [M/H]$\, \approx-1$ and 0 (Cooksey et al. 2008, Tripp et
al. 2006, Aracil et al. 2006, Lehner et al. 2006, Prochaska et
al. 2004), although lower metallicities are found as well (e.g.,
Stocke et al. 2007 and previous references).  The median metallicity
in the higher redshift $z\approx2-3$ LAF is [M/H]$\, \la -2$ (e.g.,
Aguirre et al. 2008; Simcoe, Sargent, \& Rauch 2004; Schaye et
al. 2003), although Simcoe et al. (2006) find that the small fraction
of the IGM nearest to galaxies is at metallicities close to those
found at low-\z.  Generally it is thought that significant pollution
of the average diffuse IGM has to have occurred since $z\sim3$ (Stocke
et al. 2007, \dave\ \& Oppenheimer 2007, Tumlinson \& Fang 2005).

Generally speaking there are three uncertainties associated with the
ionization of the $z\approx0.495$ system toward \pks\ (and others): 1)
Is the source of the ionization collisional ionization,
photoionization, or a mixture of both? 2) If photoionization is
important, what is the nature of the ionizing UVB spectrum? and 3) If
collisions are important, what assumptions are appropriate (e.g., CIE
versus various NEQ scenarios) and are the models sufficiently
sophisticated?  Of course, the latter concern of the level of
sophistication in the models is appropriate for all processes.  The
$z\approx0.495$ absorber toward \pks\ has characteristics consistent
with pure photoionization by the UVB, perhaps including contributions
radiation that has leaked from galaxies (Haardt \& Madau 2001).
Models of pure collisional ionization do not reproduce the observed
ionic ratios in this absorber.  Multiphase models that include the
effects of ionization via both photons and electron collisions are
also broadly consistent with the observed ionic ratios in this
absorber.  Even in these models, however, the majority of the gas is
likely warm photoionized material.

The choice of one or the other of the ionizing backgrounds considered
here is an important one, as they give significantly different
metallicities for the absorber, ranging from [O/H]$\, \approx -0.62$
for the QSOs+galaxies UVB to $\approx-0.15$ for the QSOs only
spectrum.  Thus, the choice of one of two reasonable UVB spectra (and
other UVB prescriptions exist) gives metallicities discrepant by
$\sim0.5$ dex, or a factor of three.  The difference comes about
because the QSOs+galaxies model requires more H-ionizing photons to
match the \ovi\ column.  Thus, for the same \oiv /\ovi\ or \ovi
/\ciii\ ratio, the neutral fraction, $x(\mbox{\HI})$, is lower in the
QSOs+galaxies model, implying a larger total H column density for the
same metal ion columns.  This difference can have significant
consequences for understanding the enrichment of the low-\z\ IGM and
the low-\z\ baryon budget.

Thus there is a large systematic uncertainty associated with the
choice of ionizing background model (e.g., see discussions in Tripp et
al. 2008, Schaye et al. 2003, Giroux \& Shull 1997), one that is not
always explicitly considered in the study of the low-\z\ IGM.  There
are several plausible UVB models available, such as those dominated by
AGN or QSOs with varying spectral slopes (e.g., Telfer et al. 2002,
Mathews \& Ferland 1987) or those containing contributions from
ionizing radiation that has escaped galaxies (e.g., the Haardt \&
Madau spectrum adopted here).

Determining the spectral shape of the UVB is an on-going challenge.
We have shown that the $z\approx 0.495$ absorber toward \pks\ can
plausibly be ionized via photoionization by a UVB including a
contribution from galaxies with $f_{\rm esc}\sim0.1$.  There is some
evidence for a significant contribution to the UVB from galaxies.
Faucher-Gigu\`{e}re et al. (2008) have recently presented evidence
based on the ionization rate of the $z\sim2-4$ LAF that galaxies may
dominate the ionizing UVB over much of that range, although the extent
to which they contribute at low redshift is less well constrained.

The UVB may also be spatially variable, depending on the proximity of
the absorber to galaxies or AGN; thus, the specific shape adopted may
depend on unknown properties of the absorber.  Reimers et al. (2006)
have argued such variation is likely for $z\approx2$ \ovi\ absorbers,
and there is strong evidence for a varying UVB spectral shape from the
\heii\ forest at $z\approx 2.5$ (e.g., Shull et al. 2004, Kriss et
al. 2001).  Oppenheimer \& \dave\ (2009) have also argued for a
spatially-variable ionizing background in order to explain the broad
range of observed \ovi / \HI\ ratios (Tripp et al. 2008, Thom \& Chen
2008b) compared with their simulations.  If the UVB can have
significant contributions from relatively local sources, leaky low
luminosity galaxies ($L\la0.1L_*$) are mostly likely to influence the
ionization of the present absorber given the lack of high luminosity
galaxies with low impact parameter to the QSO sight line (Prochaska et
al. 2006; see \S \ref{sec:discussionsub1}).

Our data do not provide sufficient diagnostics to distinguish between
the two UVB models investigated here.  One would ideally like to use a
large number of ionic measurements to constrain the contribution of
galaxies to the UVB and the spectral slope of the higher energy
component dominated by QSOs.  A larger sample of low-\z\ absorbers for
which several ionization stages of oxygen can be measured will be
collected with the upcoming Cosmic Origins Spectrograph; these may be
useful in discriminating UVB spectral shapes.  The transitions from
\oii, \oiii, \oiv, and \ov\ are shifted into the COS bandpass for
redshifts $z\approx0.38$, 0.38, 0.46, and 0.83, respectively.  Thus,
one will have access to \oii\ -- \oiv+\ovi\ for $z\sim0.5$ absorbers with
COS.  As discussed below, the addition of \civ\ observations so that
multiple ionization stages of both C (\cii, \ciii, and \civ) and
O are measured may also help distinguish various models for the UVB.
Unfortunately, \civ\ will reside in the NUV channel of COS for
absorbers $z\ga0.12$; the NUV channel is significantly less efficient
than the FUV channel (both in throughput and spectral coverage per
exposure).  Until better constraints on the UVB are available, it
seems important that elemental abundance studies of low-\z\ IGM
absorbers consider variations in the assumed energy distribution of
their adopted UVB when deriving metallicities.

Based on their measurements of \HI, \ciii, and \ovi, Danforth et
al. (2006) argued that no single phase model can appropriately match
their observations.  Danforth et al. considered only an AGN-dominated
spectrum (though not the one adopted here).  We have shown that a
single phase photoionization model can match a broad suite of ionic
column densities in the present absorber.  Other studies have shown
that pure photoionization can be made consistent with the ionization
characteristics of a significant fraction of the low-\z\ \ovi\
absorbers.  For example, Tripp et al. (2008) showed that the \ovi
/\HI\ and \ovi /\ciii\ column density ratio in a large sample of
absorbers could be explained by simple photoionization models.  In
their simplest models, they assumed the absorbers have densities
consistent with hydrostatic equilibrium in the low-\z\ IGM (following
Schaye 2001) and could match the full range of \ovi /\HI\ ratios if
admitting a significant dispersion of densities about the mean, a
range of metallicities, and/or spatially varying UVB intensity and
shape, none of which is ruled out at this point.  Matching the \ciii
/\ovi\ ratio in the few absorbers for which Tripp et al. had
measurements of that ratio required a departure from the hydrostatic
assumption, but their assumptions were still reasonable in the context
of the IGM.

Thus, Danforth et al. (2006) conclude photoionization is unlikely on
the basis of a photoionization model matching an ensemble of absorbers
with a small number of ionic measurements to ionization models that
use only one UVB that photoionization.  Meanwhile, Tripp et al. (2008)
come to opposing conclusions, finding that photoionization may explain
the conditions in an ensemble of aborbers by comparing a wide range of
models using different UVB and density assumptions to an ensemble of
\ovi\ and \ciii\ measurements.  It seems very difficult, given the
work on the present absorber, to rule out either photoionization or
collisional ionization in such a broad range of absorbers, especially
given the small number of ions typically measured and the large number
of potential model assumptions.

Danforth \& Shull (2008) have presented measurements for a broader
range of ions, including \ciii, \civ, \nv, \ovi, \siiii, and \siiv, in
a survey of low-\z\ absorbers.  On the basis of an intercomparison of
the various ionic ratios, these authors argue that \ovi\ and \nv\ are
reliable tracers of the WHIM, while \civ\ may arise from either
collisionally-ionized or photoionized material.  The rest of the ions
they argue are relatively good tracers of photoionized matter.  We
feel that detailed models of the individual absorbers are likely
required to determine whether this is truly the case (e.g., Cooksey et
al. 2008; Prochaska et al. 2004), especially given the range of
potential ionizing backgrounds and the potential for mixed ionization
processes (photoionization and collisional ionization).  In
particular, we do not find the argument that the slopes of the
frequency distributions in \ovi\ and \ciii\ differ a strong argument
for the dominance of collisionally-ionized \ovi, since a distribution
of densities, ionization parameters, and local shape of the UVB may
provide for much of the difference.  We are not arguing that all of
the \ovi\ is photoionized, but rather that it may be difficult to tell
even in absorbers with observations of a wide range of ions.

Thus, while there is strong evidence that some of the \ovi\ absorbers
are associated with hot, collisionally ionized gas (e.g., Richter et
al. 2004, Savage et al. 2005, Tripp et al. 2001), a significant
fraction of these systems may be strongly influenced by
photoionization, like the present system, which requires $\sim25\%
$--100\%\ of the \ovi to be photoionized.  These populations are not
exclusive given the multiphase nature of some absorbers and the
possibility that some trace moderate temperature ($T<10^5$ K)
photoionized gas that is mildly shock heated or that has cooled from
higher temperatures (e.g., models discussed by Oppenheimer \& \dave\
2009, Kang et al. 2005, Furlanetto et al. 2004 and others).  Tripp et
al. (2008) present temperature estimates for a number of the
well-aligned systems in their sample.  For those abosrbers with
$T<10^5$ K, a fair number seem to have temperatures somewhat higher
than one would expect from pure photoionization, suggesting an extra
source of heating that would likely impact the ionization as well.
These measurements suggest a mixture of processes could be at work in
some absorbers, although photoionization is likely a strong component
for a large fraction of the aligned systems.  It is not yet clear what
the mixture is like in the multiphase absorbers so categorized due to
differences in the \HI\ and \ovi\ profiles.

Prochaska et al. (2004) have noted an apparent anticorrelation between
the ionization parameter for the Cloudy photoionization model best
fitting the metal column densities and the \HI\ column of an absorber,
a result also seen by Lehner et al. (2006) in a larger number of
absorbers along several sight lines.  This can be understood in part
if the \HI\ column is a rough measure of physical density of the
absorber (e.g., Schaye 2001; see discussion in Prochaska et al. 2004).
The relationship seen in these earlier works will certainly depend on
the adopted ionizing spectrum, since the two UVB spectra investigated
here give significantly different ionization parameters.  However,
while those studies adopt a Haardt \& Madau QSOs-only spectrum, if
this result is directly tied to the physical density of the
absorption, the relationship should still hold if galaxies contribute
to the ionization.  While it has been suggested this correlation
implies the absorbers in those studies are likely to be photoionized,
our multiphase models show that a mixture of ionization mechanisms can
not only give similar ionization parameters for the photoionized
component, but in some cases a photoionization only and multiphase
ionization model cannot be distinguished.  Thus, the correlation
likely does not rule out a contribution to those absorbers from
collisional ionization, although it may suggest photoionization
dominates the ionization, as in the absorber studied in this work.

An important remaining question is how one might distinguish
multiphase models such as those with results summarized in \S
\ref{sec:multiphase} from the pure photoionization models of \S
\ref{sec:photoionization}.  While the aforementioned ionization
parameter-\HI\ column relationship may not discriminate between these
models, there are several promising options.  For the current
absorber, \civ\ has significant power to discriminate between these
classes of models.  The ratios \civ /\ciii\ and \civ /\ovi\ are
significantly different for photoionization models compared with the
multiphase models due to the strong variations in these ratios as a
function of temperature in the collisionally ionized gas.  At low
temperatures, the \civ /\ciii\ ratio in our multiphase models is
nearly the same as the QSOs+galaxies UVB photoionization model.
However, for temperatures $\log T \ga 5.3$, this ratio is $\sim2$ to
15 times higher for the multiphase models than the photoionization
models.  Conversely, the \civ /\ovi\ ratio is within a factor of two
for both models at $\log T \ga 5.3$, while at lower temperatures the
multiphase models predict a \civ /\ovi\ ratio factors of $\sim$4 to
$>$1000 higher than the pure photoionization models (which give
$N(\mbox{\civ})/N(\mbox{\ovi}) \sim 0.6$ to 0.2 for the QSOs and
QSOs+galaxies UVB, respectively).  This large difference in models at
low temperatures is due to the strong fall off in \ovi\ ionization
fraction with decreasing temperature over this range. These large
differences for models in which a small fraction of the gas is
associated with the collisionally-ionized phase is due to the very
large (or small) ratios of \civ\ compared with the other two ions in
collisionally ionized gas.  For example, at $\log T = 5.45$, the \civ
/\ciii\ ratio in CIE is expected to be $\sim$60 compared with $\sim$2
in the QSOs+galaxies photoionization model.  Thus, even though the
collisionally ionized phase is limited to $\la50\%$ of the total
hydrogen at this temperature (see Figure \ref{fig:multiphase}), it
contributes $>90\%$ of the \civ.

Danforth \& Shull (2008) studied the integrated column densities of
\ciii, \civ, and \ovi\ for a number of absorbers, finding \ciii /\civ\
ratios of $\sim1$ and \civ /\ovi\ ratios consistent with $\sim0.25$
(both with a quite large scatter).  Both of these ratios are
consistent with our photoionization models, though it may be dangerous
to draw conclusions based on these two average ratios, which rely on
different samples of absorbers, without attempting models of the
ionization to match the full range of ionization states in each
absorber.  

Indeed, Danforth \& Shull conclude that the \ovi\ mostly arises in
collisionally-ionized gas on the basis of their comparisons.  Danforth
\& Shull note that the correlation of \civ\ with both \ovi\ and \ciii\
is not particularly strong, and they use this to argue that the
observed \civ\ in their sample may arise from both shock-heated and
photoionized gas (traced by \ovi\ and \ciii, respectively, in their
argument).  A significant population of multiphase absorbers where the
mixture of phases varies could presumably also provide for such a lack
of correlation.

Future observations with the Cosmic Origins Spectrograph (COS) on
board \hst\ should provide significantly higher signal-to-noise ratio
observations of QSOs than have been possible with STIS, albeit at
lower spectral resolution.  This can provide better limits and
detections of weak absorption lines for some sight lines with quite
low-S/N STIS observations, such as the \pks\ sight line, allowing
stronger constraints on the models.  In the present absorber, for
example, \oiii\ already provides important constraints on the
single-phase collisional ionization models.  Better observations of
this transition will begin to limit the photoionization and multiphase
models more severely. 

The metallicity of this absorber is quite high for both UVB spectra
adopted here.  There are significant systematic uncertainties
associated with our estimates, but [O/H]$\, \sim-0.6$ or $-0.15$ are
both high compared with the canonical mean of [O/H]$\, -1$.  The high
metallicity and great strength of the \ovi\ absorption both favor an
origin of this absorber near to a galaxy, although the \HI\ column and
estimated densities are not large.  Given the discussion in the
previous section, if this absorber were associated with a galaxy, it
would likely be a sub-$L_*$ system.

\section{Summary}
\label{sec:summary}

We have presented \fuse\ and \hst /STIS ultraviolet absorption line
observations of the $z = 0.495096$ absorber toward the QSO \pks.  We
have measured the column densities of \HI, \ciii, \niv, \oiv, \ov, and
\ovi\ and placed upper limits on the column densities of another seven
ions.  We use these measurements to study the ionization processes at
work in this absorber and estimate its metallicity.

The most important results of our work are as follows.

\begin{enumerate}

\item This absorber shows very strong \ovi\ absorption with no
  detectable \neviii\ or \mgx.  There is no direct evidence for a
  strong component of gas with temperatures $T\ga5\times10^5$ K as
  expected if it traces the largest mass of missing baryons in a WHIM.
  However, the limits to the amount of material depend strongly on the
  temperature (see Figure \ref{fig:whimlimits}).  This system would
  not likely be detectable in X-ray absorption.

\item We have modeled the broad range of ions covered by our data
  using a number of collisional ionization and photoionizations
  models.  This absorber can be modelled as purely photoionized gas
  with [O/H]$\, \sim -0.6$ if the ionizing UVB includes a significant
  contribution from photons associated with star forming galaxies and
  [O/H]$\, \sim -0.15$ if the UVB does not include a significant
  contribution from galaxies.  These metallicities are robust to the
  inclusion of a collisionally ionized phase.
  
\item The ionization of the absorber may also be well described by
  multiphase models in which the photoionized gas in the absorber is
  complemented by a contribution from hot ($T\ga3\times10^5$ K),
  collisionally ionized gas.  The strong \ovi\ in this system may
  trace a photoionized phase, a collisionally-ionized phase, or a
  mixture of both.  In this case, the metallicities are still
  consistent with those from a pure photoionization model.

\item The uncertainties in the spectral shape of the ionizing UVB give
  rise to significant systematic uncertainties in the derived
  metallicity of the absorber, of order 0.5 dex or a factor of 3 in
  this case.  Such uncertainties should be fully considered when
  studying the metallicity distribution, ionization, and baryon
  content derived from metal lines of the low-\z\ IGM.

\end{enumerate}

\section*{Acknowledgments}

We thank A.J. Fox, J.M. Shull, and T.M. Tripp for comments that
improved our work.  JCH and JXP recognize support from NASA grant
NAG5-12345.  JXP acknowledges support from NASA grant NAG5-12743.  JCH
and NL also recognize support from NASA grant NNX08AJ51G.



\bsp

\label{lastpage}

\end{document}